\documentclass[twocolumn,showpacs,amsmath,amssymb,nofootinbib,pra,superscriptaddress]{revtex4-1}

\usepackage{comment}

\usepackage{xcolor}

\usepackage{graphicx}
\usepackage{epsfig}
\usepackage{dcolumn}
\usepackage{bm}
\usepackage{braket}
\usepackage{amsmath}
\usepackage{mathtools}
\usepackage{color}
\usepackage{hyperref}
\newcommand{\abs}[1]{\left| #1 \right|} 

\usepackage{hyperref}
\usepackage{multirow}
\begin{document}

\title{Induced correlations between impurities in a one-dimensional quenched Bose gas}

\author{S. I.~\surname{Mistakidis}}
\affiliation{Center for Optical Quantum Technologies, Department of
Physics, University of Hamburg,
Luruper Chaussee 149, 22761 Hamburg Germany} 

\author{A.~G.~\surname{Volosniev}}
\affiliation{Institut f{\"u}r Kernphysik, Technische Universit{\"a}t Darmstadt, 64289 Darmstadt, Germany}
\affiliation{IST Austria, am Campus 1, 3400 Klosterneuburg, Austria}
\author{P. Schmelcher}
\affiliation{Center for Optical Quantum Technologies, Department of
Physics, University of Hamburg,
Luruper Chaussee 149, 22761 Hamburg Germany} 
\affiliation{The Hamburg
Centre for Ultrafast Imaging,
Universit\"{a}t Hamburg, Luruper Chaussee 149, 22761 Hamburg, Germany}

\begin{abstract}
We explore the time evolution of two impurities in a trapped one-dimensional Bose gas that follows a change of the boson-impurity interaction. 
We study the induced impurity-impurity interactions and their effect on the quench dynamics.  In particular, we report on the size of the impurity cloud, the impurity-impurity entanglement, and the impurity-impurity correlation function. 
The presented numerical simulations are based upon the variational multilayer multiconfiguration time-dependent Hartree method for bosons.
To analyze and quantify induced impurity-impurity correlations, we employ an effective two-body Hamiltonian with a contact interaction. We show that the effective model consistent with the mean-field attraction of two heavy impurities explains qualitatively our results for weak interactions.  
 Our findings suggest that the quench dynamics in cold-atom systems can be a tool for studying impurity-impurity correlations.   
\end{abstract}

\maketitle

\section{Introduction}
Nature provides us with a multitude of highly imbalanced two-component systems in which the first component contains many more particles than the second one. To understand certain properties of these systems, it is reasonable to reduce the second component to a single particle, and analyze a model problem of an environment with an impurity. Complementary studies then should address the role of the impurity-impurity interactions by, for example, investigating systems with two impurities. 
This two-stage approach has been exploited in classic many-body problems. For example, it was used to investigate electrons in crystals leading to, e.g., polarons, F-centers~\cite{pekar1963, alexandrov1995}, and
$^3$He particles in $^4$He~\cite{mahan2000, baym2008}. 

Ultracold atoms allow one to engineer and explore systems with population imbalance~\cite{zwierlein2009,nascimbene2009,chevy2010,catani2012,widera2012,grimm2012,koschorreck2012,arlt2016,
jin2016,scazza2017,zwielein2019}, and the corresponding impurity-impurity induced interactions~\cite{chin2019,edri2019}. The cold-atom setups are tunable and adjustable to the physics of interest, and as such they provide a testground for the concepts of dressed particles and of induced impurity-impurity interactions.  It is important to find parameter regimes for which the latter of the two concepts has a visible impact, since the corresponding signatures are often difficult to observe. In this paper, we investigate the possibility of detecting impurity-impurity interactions in weakly-interacting one-dimensional Bose gases. To this end, we study the quench dynamics that follows a rapid change of the boson-impurity interaction strength. 

Let us briefly address what is known about impurity-impurity interactions induced by a weakly-interacting Bose gas in one spatial dimension, see Refs.~\cite{naidon2018,bruun2018,bruun2018a,pastukhov2019} for recent studies on two impurities in cold three-dimensional Bose gases, and Refs.~\cite{flicker1967, fukuhara2013, schecter2014, huber2019, pasek2019} for a discussion of impurities in strongly-interacting Bose gases.  Homogeneous systems with weakly-interacting and heavy impurities are well-understood by now. The interaction at short distances follows an attractive exponential function~\cite{recati2005, bruderer2007,dehkharghani2018,reichert2018}. At long distances, quantum fluctuations lead to a power-law decay of interactions~\cite{schecter2014,reichert2018,pavlov2018}. One can argue that weakly-interacting mobile impurities interact similarly~\cite{schecter2014}. To observe induced interactions in cold-atom systems, one should understand finite-size as well as the trap effects. A number of works studied these effects using time-independent models at various parameter regimes~\cite{dehkharghani2018, keiler2018, chen2018, lewenstein2019,pasek2019}. 

Our work is in line with the previous studies on induced interactions in harmonically trapped static systems. Moreover, it extends them by investigating a corresponding dynamical problem. There are two main motivations for the present work. {\it First}, we aim at understanding for which parameters and observables an effective interaction derived for a homogeneous medium can be used 
to adequately describe the dynamics in a harmonic trap. 
Note that the induced interaction in a harmonic trap is not Galilean invariant~\cite{dehkharghani2018}, which complicates the use of a Galilean invariant effective potential characteristic for a homogeneous environment. As we show, this complication can be avoided by considering weakly-interacting systems.  {\it Second}, we want to estimate the effect of the induced interactions on the dynamics. 

In this paper, we introduce an effective model for impurity-impurity interactions and compare it with the multilayer multiconfiguration time-dependent Hartree method for bosons (ML-MCTDHB)~\cite{kronke2013, cao2013, mistakidis2017, Mistakidis_cor, Katsimiga_DBs, Koutentakis_FF, expansion}.
It is shown that a two-body effective model with a zero-range potential is able to explain qualitatively the time evolution of two impurities that interact weakly with a trapped Bose gas. For moderate interactions, we observe that the impurity-impurity interaction cannot be modelled using a delta function. Our findings demonstrate that even weak interactions noticeably affect the dynamics. We conclude that the quench dynamics may be used to observe induced interactions experimentally. 

The paper is organized as follows. Section~\ref{sec:ham} formulates the time-dependent problem we consider: a trapped Bose gas with two impurity atoms being initially in the ground state for vanishing boson-impurity interactions. The time dynamics is then initiated by a sudden change of the boson-impurity interaction strength. Section~\ref{sec:eff_ham} introduces an effective two-body Hamiltonian
that we use to study the time evolution of the impurity-impurity subsystem. Section~\ref{sec:res} presents the time evolution of the size of the impurity cloud, the entropy and the two-body coherence function. In that section, we compare the numerical data with the effective model from Sec.~\ref{sec:eff_ham}. Section~\ref{sec:summary} summarizes our results and provides an outlook. Three appendices contain the technical details of our work. Appendix~\ref{methodology} reviews the ML-MCTDHB method. Appendix~\ref{sec:convergence} discusses the accuracy of the numerical data, and Appendix~\ref{sec:app_a} elaborates on the effective two-body Hamiltonian.

\section{Hamiltonian}
\label{sec:ham}

We consider two bosonic impurities in a system of $N$ bosons. The corresponding Hamiltonian reads
\begin{align}
H&=\sum_{i=1}^N h(x_i)+\sum_{j=1}^2 h(y_j)+g_{BB}\sum_{i>j}\delta(x_i-x_j) \nonumber \\&+g_{IB}\sum_{i=1}^{N}\sum_{j=1}^{2}\delta(x_i-y_j)+g_{II}\delta(y_1-y_2),
\label{eq:ham}
\end{align}
where $y_j$ is the position of the $j$th impurity, and $x_i$ is the position of the $i$th boson. The parameters $g_{IB}$, $g_{BB}$ and $g_{II}$ determine the strengths of the zero-range interactions.
Without interactions all particles are described by the identical one-body Hamiltonians
$h(z)=-\frac{\hbar^2}{2m}\frac{\partial^2}{\partial z^2}+\frac{k z^2}{2}$. 
Here $m$ is the mass of a particle, $k= m\Omega^2$ is the spring constant, and $\Omega$ is the frequency of the external harmonic trap. The ground state properties of the Hamiltonian~(\ref{eq:ham}) were studied using a variational wave function in Ref.~\cite{dehkharghani2018}, see also~\cite{mistakidis2019polarons}. In this work, we focus on the dynamics following a sudden change of $g_{IB}$ ($g_{IB}=0$ at $t=0$, $g_{IB}\neq 0$ at $t>0$) assuming that the system is in the ground state at $t=0$. 
The time evolution of the system obeys the time-dependent Schr{\"o}dinger equation 
\begin{equation}
i\hbar \frac{\partial \psi(y_1,y_2,x_1,...,x_N;t)}{\partial t}=H\psi(y_1,y_2,x_1,...,x_N;t),
\label{eq:schr_equation}
\end{equation} 
where the many-body wave function, $\psi$, satisfies the initial condition:
$\psi(t=0)$ is the ground state of $H(g_{IB}=0)$. We tackle Eq.~(\ref{eq:schr_equation}) using the ML-MCTDHB method~\cite{kronke2013, cao2013, mistakidis2017}, see Appendices~\ref{methodology} and~\ref{sec:convergence} for a brief description of the method and the accuracy of the calculations. ML-MCTDHB is a variational method for the many-body dynamics that allows us to numerically study beyond mean-field correlations between particles. In the ML-MCTDHB approach, the many-body wavefunction is expanded in a time-dependent and variationally 
optimized basis that spans the most important part of the Hilbert space and disregards the remainder. To construct this basis, we first write the many-body wave function as a truncated Schmidt decomposition using $D$ species functions for each component, see also Eq.~(\ref{Eq:WF}) in Appendix~\ref{methodology}. 
As a next step, we expand the aforementioned species functions in a basis of $d_B$ ($d_I$) single-particle functions for the bosonic medium (the impurities), see also Eq.~(\ref{Eq:SPF}) in Appendix~\ref{methodology}. Then, a variational principle leads to a set of nonlinear integro-differential equations. For convenience, the single-particle functions are expanded with respect to a time-independent primitive basis. In our case, this primitive basis is given by a sine discrete variable representation of a one-body space with hard-wall boundary conditions at both edges of the numerical grid. The grid consists of 500 points. 
 
Our system constitutes of a Bose gas with many weakly interacting particles and two imputity atoms. A large number of bosons can be adequately described using a small number of single-particle functions. In our calculations, we established that $d_B=3$ captures the dynamics well. 
The number of impurities is small and we need to use more single-particle functions, $d_I\simeq 8$, to obtain accurate results.
It is worthwhile noting that the variational nature of ML-MCTDHB allows us to estimate its accuracy by varying $D, d_I$ and $d_B$. 
For details on the precision of our numerical results see Appendix~\ref{sec:convergence}. 

\begin{figure}
\centerline{\includegraphics[scale=0.31]{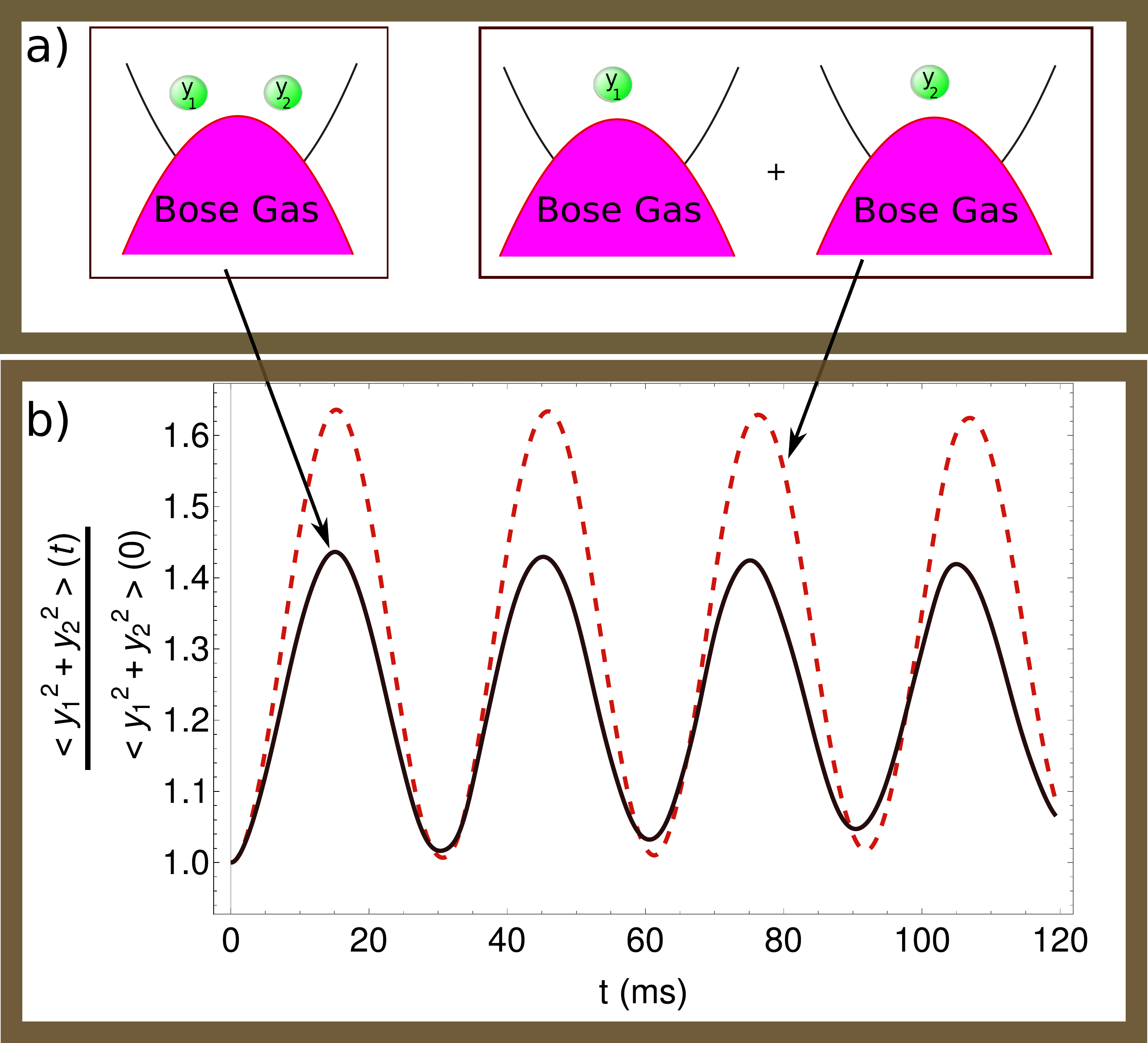}}
\caption{Panel {\bf a)} illustrates the two systems of interest: (left) a Bose gas with two impurity atoms, and (right) two Bose gases, each with a single impurity atom. Panel {\bf b)} shows the time evolution of $\langle y_1^2+y_2^2\rangle(t) /\langle y_1^2+y_2^2\rangle(0)$ after a rapid change of the boson-impurity interaction strengths.  The (red) dashed curve describes the impurities in different traps~\cite{mistakidis2018}. The (black) solid curve presents the dynamics of two impurities in the same trap. There is no free space impurity-impurity interaction, $g_{II}=0$, and we interpret the difference between the two curves as a manifestation of attractive induced impurity-impurity interactions, see the text for details.  The final interaction strength is $g_{IB}(t>0)/g_{BB}$=0.3, all other parameters (e.g., the number of bosons, $N=100$) are common for all numerical data and are given in the text.}
\label{fig:fig1}
\end{figure}

The dynamics of a single impurity has already been explored using the ML-MCTDHB~\cite{mistakidis2018,mistakidis2019,mistakidis2019a}. In this paper, we seek insight into induced impurity-impurity correlations. We compare the dynamics of two impurities in a Bose gas to that of a single impurity presented in Ref.~\cite{mistakidis2018}, see also Fig.~\ref{fig:fig1}. Therefore, for our numerical simulations, we use the parameters of Ref.~\cite{mistakidis2018}, namely: $N=100, \Omega=2\pi\times 20 Hz, g_{BB}=10^{-37}J m$, and $m=m(^{87}\mathrm{Rb})$. To give a sense of the size of the Bose gas for these parameters, we note that the corresponding Thomas-Fermi radius is approximately $19 \mu$m.

%%%%%%%%%%%%%%%%%%%%%%%%%%%%%%%%%%%%%%%%%%%%%%%%%%%%%%%%%%%%%%%%%%%%%%%%%%%%%%%%%%%%%%%%%%%%%%%%%%%%%%%%%%%%%%%%%%%%%%%
%%%%%%%%%%%%%%%%%%%%%%%%%%%%%%%%%%%%%%%%%%%%%%%%%%%%%%%%%%%%%%%%%%%%%%%%%%%%%%%%%%%%%%%%%%%%%%%%%%%%%%%%%%%%%%%%%%%%%%%

To illustrate the effect of induced interactions in our model, we
determine the square of the size of the cloud of two impurities placed in a Bose gas:
\begin{equation}
\langle y^2_1 + y^2_2 \rangle_{MB}=\int \mathrm{d}y_1\mathrm{d}y_2\mathrm{d}x_1 \cdots \mathrm{d}x_N (y_1^2+y_2^2)|\Psi_{MB}|^2,
\label{eq:size_y1y2}
\end{equation} 
where $\Psi_{MB}$  is obtained using the ML-MCTDHB method (see Appendix~\ref{methodology}). 
To extract information about induced impurity-impurity correlations, the quantity $\langle y^2_1 + y^2_2 \rangle_{MB}$ is 
compared to the square of the size of the cloud of two impurities in two separate Bose gases [see Fig.~\ref{fig:fig1}{\bf a)}]:
\begin{equation}
\langle y^2_1+y^2_2 \rangle_{1,MB} = 2\int \mathrm{d}y\mathrm{d}x_1 \cdots \mathrm{d}x_N y^2|\psi_1|^2,
\label{eq:size_y1y1}
\end{equation}
where $\psi_1$ is the wave function of a Bose gas with only one impurity atom (see Ref.~\cite{mistakidis2018}). We present our results in Fig.~\ref{fig:fig1} {\bf b)} for $g_{II}=0$ and $g_{IB}(t>0)/g_{BB}=0.3$. Figure~\ref{fig:fig1}~{\bf b)} demonstrates that following the sudden change of $g_{IB}$ the square of the size of the impurity cloud determined by Eq.~(\ref{eq:size_y1y2}) [correlated impurities] features a breathing oscillation motion similar to that of Eq.~(\ref{eq:size_y1y1}) [non-correlated impurities]. However, $\langle y^2_1+y^2_2 \rangle_{1,MB}(t)$ is noticeably different from $\langle y^2_1 + y^2_2 \rangle_{MB}(t)$ [note that $g_{IB}(t=0)=0$, hence $\langle y^2_1 + y^2_2 \rangle_{MB}(t=0)=\langle y^2_1+y^2_2 \rangle_{1,MB}(t=0)$]. To interpret Fig.~\ref{fig:fig1}, we introduce in the next section an effective description based on dressed impurities. According to that model, the difference between the curves in Fig.~\ref{fig:fig1}~{\bf b)} is a manifestation of induced impurity-impurity interactions. Below, we discuss this effective description, which allows us to visualize and quantify the induced impurity-impurity correlations. This description is a useful tool that gives insight into systems with multiple impurities and does not rely on elaborate many-body simulations.

\section{Effective Two-Body Hamiltonian}
\label{sec:eff_ham}

The simplest phenomenological approach to a single impurity in a Bose gas is the use of an 
effective one-body Hamiltonian. In a homogeneous case, such a Hamiltonian is written as follows
\begin{equation} 
h_1=\epsilon-\frac{\hbar^2}{2m_{\mathrm{eff}}}\frac{\partial^2}{\partial y^2},
\end{equation}
where $\epsilon(g_{II},g_{IB},\rho)$ and $m_{\mathrm{eff}}(g_{II},g_{IB},\rho)$
are, respectively, the self-energy and the effective mass of the dressed impurity atom~\cite{catani2012,kain2016, parisi2017,volosniev2017,grusdt2017,pastukhov2017,mistakidis2018}, here $\rho$ is the density of the homogeneous Bose gas. 
In this paper, we are interested in impurity-impurity correlations in a trap where the momenta of the impurities are non-vanishing. Therefore, the emergent correlations between impurities can be considered as a perturbation, which can be accounted for by some potential $V$ that mimics the relevant low-energy transitions between the states of $h_1$. We choose the two-body effective Hamiltonian for a homogeneous case, $k=0$, as
\begin{equation}
H^{k=0}_{\mathrm{eff}}=2\epsilon-\frac{\hbar^2}{2m_{\mathrm{eff}}}\frac{\partial^2}{\partial y_1^2}-\frac{\hbar^2}{2m_{\mathrm{eff}}}\frac{\partial^2}{\partial y_2^2}
+ V(y_1-y_2).
\label{eq:eff_k=0}
\end{equation}
Our focus are weak correlations for which the exact shape of $V$ is not important. For simplicity,
we assume that 
\begin{equation}
V(y_1-y_2)=[g_{II}^i(\rho,g_{IB},g_{BB})+g_{II}]\delta(y_1-y_2),
\end{equation}
where the parameter $g_{II}^i$ defines the strength of the induced interactions.

{\it Trapped case.} The Hamiltonian~(\ref{eq:eff_k=0}) is not applicable to the experiments with trapped cold atoms. 
To extend Eq.~(\ref{eq:eff_k=0}) to a harmonically trapped case, we assume 
that the density of the Bose gas varies slowly on the length scale given by the healing length. In this case, we can rely on the local density approximation to derive the effective two-body Hamiltonian for a trapped system
\begin{align}
H^{k}_{\mathrm{eff}}=\epsilon(y_1)+\epsilon(y_2)-\frac{\hbar^2}{2m_{\mathrm{eff}}(y_1)}\frac{\partial^2}{\partial y_1^2}
-\frac{\hbar^2}{2m_{\mathrm{eff}}(y_2)}\frac{\partial^2}{\partial y_2^2} \nonumber \\ +\frac{k y_1^2}{2}+\frac{k y_2^2}{2}+[g_{II}^i(\rho(y_1),g_{IB},g_{BB})+g_{II}]\delta(y_1-y_2),
\label{eq:eff_ham_k}
\end{align}
where we arbitrarily choose to write $g_{II}^i(\rho(y_1),g_{IB},g_{BB})$ instead of $g_{II}^i(\rho(y_2),g_{IB},g_{BB})$ -- both expressions are identical because the interaction term in Eq.~(\ref{eq:eff_ham_k}) is non-vanishing only if $y_1=y_2$.
Note that the effective interaction in Eq.~(\ref{eq:eff_ham_k}) depends, in general, on the coordinates $y_1$ and $y_2$ (cf.~Ref.~\cite{dehkharghani2018, chen2018}), and not on $y_1-y_2$ as in the homogeneous case. To simplify the description, we assume that $g_{II}^i(\rho(y_1),g_{IB},g_{BB})=g_{II}^i(\rho(0),g_{IB},g_{BB})$.
This assumption is justified only for small values of $g_{IB}$, for which the impurities move close to the center of the trap, and, therefore, experience only a weak inhomogeneity during the time evolution. 

It has been argued that $\epsilon(y_i)$ and $m_{\mathrm{eff}}(y_i)$ in Eq.~(\ref{eq:eff_ham_k}) are very sensitive to the density of bosons $\rho(y)$~\cite{mistakidis2018}. To circumvent this problem, another one-body effective Hamiltonian has been proposed~\cite{mistakidis2018}, 
\begin{equation}
h_1^{\mathrm{trap}}=-\frac{\hbar^2}{2\overline m_{\mathrm{eff}}}\frac{\partial^2}{\partial y^2}+\frac{\overline k y^2}{2},
\label{eq:h1_trap}
\end{equation} 
to analyze a single impurity atom in a trapped Bose gas.
The parameters $\overline k$ and $\overline m_{\mathrm{eff}}$ incorporate the effects of 
$m_{\mathrm{eff}}(y)$ and $\epsilon(y)$. They can be obtained by fitting to numerical and/or experimental data. We determine them as in Ref.~\cite{mistakidis2018}, i.e., we first use the ML-MCTDHB to calculate $\langle y^2 \rangle$ for a single impurity in a Bose gas, and then find the parameters in Eq.~(\ref{eq:h1_trap}) that most accurately reproduce the ML-MCTDHB results. For $g_{IB}>0$, $\overline k$ and $\overline m_{\mathrm{eff}}$ are presented in Ref.~\cite{mistakidis2018}.
Note that $h_1^{\mathrm{trap}}$ is accurate only for weak interactions. For strong interactions, the impurity minimizes the interaction with the bosons by residing at the edge of the Bose gas, which implies that a harmonic term~$\overline k y^2/2$ cannot be used in Eq.~(\ref{eq:h1_trap})~\cite{dehkharghani2015,dehkharghani2015a, dehkharghani2018,mistakidis2018,mistakidis2019}. In this paper, we focus exclusively on the regime where Eq.~(\ref{eq:h1_trap}) holds. 

We employ Eq.~(\ref{eq:h1_trap}) to mimic the effect of single-body terms in Eq.~(\ref{eq:eff_ham_k}), which leads to the two-body Hamiltonian: 
\begin{align}
\overline H_{\mathrm{eff}}&=-\frac{\hbar^2}{2\overline m_{\mathrm{eff}}}\frac{\partial^2}{\partial y_1^2}
-\frac{\hbar^2}{2\overline m_{\mathrm{eff}}}\frac{\partial^2}{\partial y_2^2} \nonumber\\ &+\frac{\overline k y_1^2}{2}+\frac{\overline k y_2^2}{2}+(g_{II}^i+g_{II}) \delta(y_1-y_2),
\label{eq:ham_eff_overline}
\end{align}
that we use to analyze the dynamics of two weakly-interacting impurities. We denote by $\Psi_{\mathrm{eff}}(y_1,y_2;t)$ the wave function that describes the time evolution governed by $\overline H_{\mathrm{eff}}$. By assumption, $\Psi_{\mathrm{eff}}(y_1,y_2;t=0)$ is the ground state of $h_1^{\mathrm{trap}}(y_1)+h_2^{\mathrm{trap}}(y_2)$. To determine $\Psi_{\mathrm{eff}}$, we first separate the relative and the center-of-mass motions. To this end, we employ the coordinates $y=(y_1-y_2)/\sqrt{2}$ and $y_{\mathrm{CM}}=(y_1+y_2)/\sqrt{2}$.
The center-of-mass dynamics is independent of the induced interactions, hence, it follows from Ref.~\cite{mistakidis2018}.
The relative motion is described by the Hamiltonian 
\begin{equation}
\overline H_{\mathrm{rel}}=-\frac{\hbar^2}{2\overline m_{\mathrm{eff}}}\frac{\partial^2}{\partial y^2}+
\frac{\overline k y^2}{2}+\frac{g_{II}^i+g_{II}}{\sqrt{2}} \delta(y).
\label{eq:eff_ham_rel}
\end{equation}
It is worthwhile noting that the wave function that describes the quench dynamics of the Hamiltonian~(\ref{eq:eff_ham_rel}) following the rapid change of parameters at $t=0$: $m\to \overline m_{\mathrm{eff}}, k\to \overline k$ and $g_{II}^i=0\to g_{II}^i\neq 0 $ for $g_{II}=0$ can be written in a closed form, see Appendix~\ref{sec:app_a}. This allows us to calculate directly all observables of interest, e.g., $\langle y_{CM}^2+y^2\rangle_{\mathrm{eff}}=\int(y_1^2+y_2^2)|\Psi_{\mathrm{eff}}(y_1,y_2;t)|^2\mathrm{d}y_1\mathrm{d}y_2$.

We derive an estimate for the parameter $g_{II}^{i}$, employing the interaction between two weakly-interacting impurities, $V_{HH}$~\cite{recati2005, bruderer2007,dehkharghani2018,reichert2018}. The potential $V_{HH}$ has a short-range (mean-field) part given by the Yukawa-type potential ($\sim e^{-r/r_0}$) and a long-range tail ($1/r^3$) due to quantum fluctuations. The long-range tail for our parameters is important at and beyond $r_{LR} \simeq 2\mu$m. We estimate this distance by comparing the value of the interaction due to the long-range part to that given by the short-range part~\cite{reichert2018}
\begin{equation}
32 \pi \rho(0)\frac{r_{LR}^3}{\xi^2}e^{-2\frac{r_{LR}}{\xi}}\simeq 1,
\end{equation}
where $\xi\simeq 0.4\mu m$ is the healing length estimated using the density of the Bose gas at the origin, $\rho(0)$. Note that $r_{LR}\simeq 5 \xi$, which implies that the effect of the long-range tail can be neglected in our analysis. In particular, we estimate the potential volume of $V_{HH}$ using only its short-range part, because a long-range part has a negligible potential volume. 
The short-range part of the potential can be derived from the density profile of a homogeneous Bose gas with a single weakly-interacting impurity~\cite{volosniev2017},
\begin{equation}
\rho(x)=\rho_0-g_{IB}\sqrt{\frac{\kappa\rho_0}{\hbar^2 g_{BB}}}e^{-2\sqrt{\frac{\kappa g_{BB}\rho_0}{\hbar^2}}|x-y_1|},
\label{eq:density_with_imp}
\end{equation}
where $\rho_0$ is the density of the Bose gas without the impurity, and $\kappa$ is the boson-impurity reduced mass. The mean-field energy of the second impurity is given by $g_{IB}\rho(y_2)$. The induced impurity-impurity interaction is determined by the second term of Eq.~(\ref{eq:density_with_imp}). The corresponding potential net volume is $\int V_{HH}(y)\mathrm{d}y\simeq -g_{IB}^2/g_{BB}$. Note that it is independent of the mass of the impurity, and thus can be derived using heavy impurities. 
Since the potential volume is the only relevant parameter for low-energy scattering, we arrive at the expression for the strength of the induced impurity-impurity interaction
\begin{equation}
g_{II}^i\simeq -\frac{g_{IB}^2}{g_{BB}}.
\label{eq:est_gIIi}
\end{equation} 

{\it Limits of applicability}. The description based on Eqs.~(\ref{eq:eff_ham_rel}) and~(\ref{eq:est_gIIi}) fully determines the dynamics of two impurities in a trapped Bose gas. Let us summarize the crucial assumptions we use to derive this description: (i) the energy scale associated with the induced impurity-impurity interaction is much smaller than all other energy scales in the problem (e.g., $\hbar\Omega$) so that this interaction can be parameterized by the delta function potential; (ii) the impurities move close to the center of the trap, such that the interaction depends only on the relative distance $y$; (iii) the interaction potential is independent of the effective mass and the spring constant, such that Eq.~(\ref{eq:est_gIIi}) can be used. One can argue that these assumptions are valid as long as $g_{IB}$ is small enough. A demonstration of this is presented in the next section, where the results of Eqs.~(\ref{eq:eff_ham_rel}) and~(\ref{eq:est_gIIi}) are compared to the results obtained within the ML-MCTDHB simulations. 

\section{Effects of the Induced Interactions on the Quench Dynamics }
\label{sec:res}

\subsection{The size of the impurity-impurity subsystem}

%%%%%%%%%%%%%%%%%%%%%%%%%%%%%%%%%%%%%%%%%%%%%%%%%%%%%%%%%%%%%%%%%%%%%%%%%%%%%%%%%%%%%%%%%%%%%%%%%%%%%%%%%%%%%%%%%%%%%%%
\begin{figure}
\centerline{\includegraphics[scale=0.28]{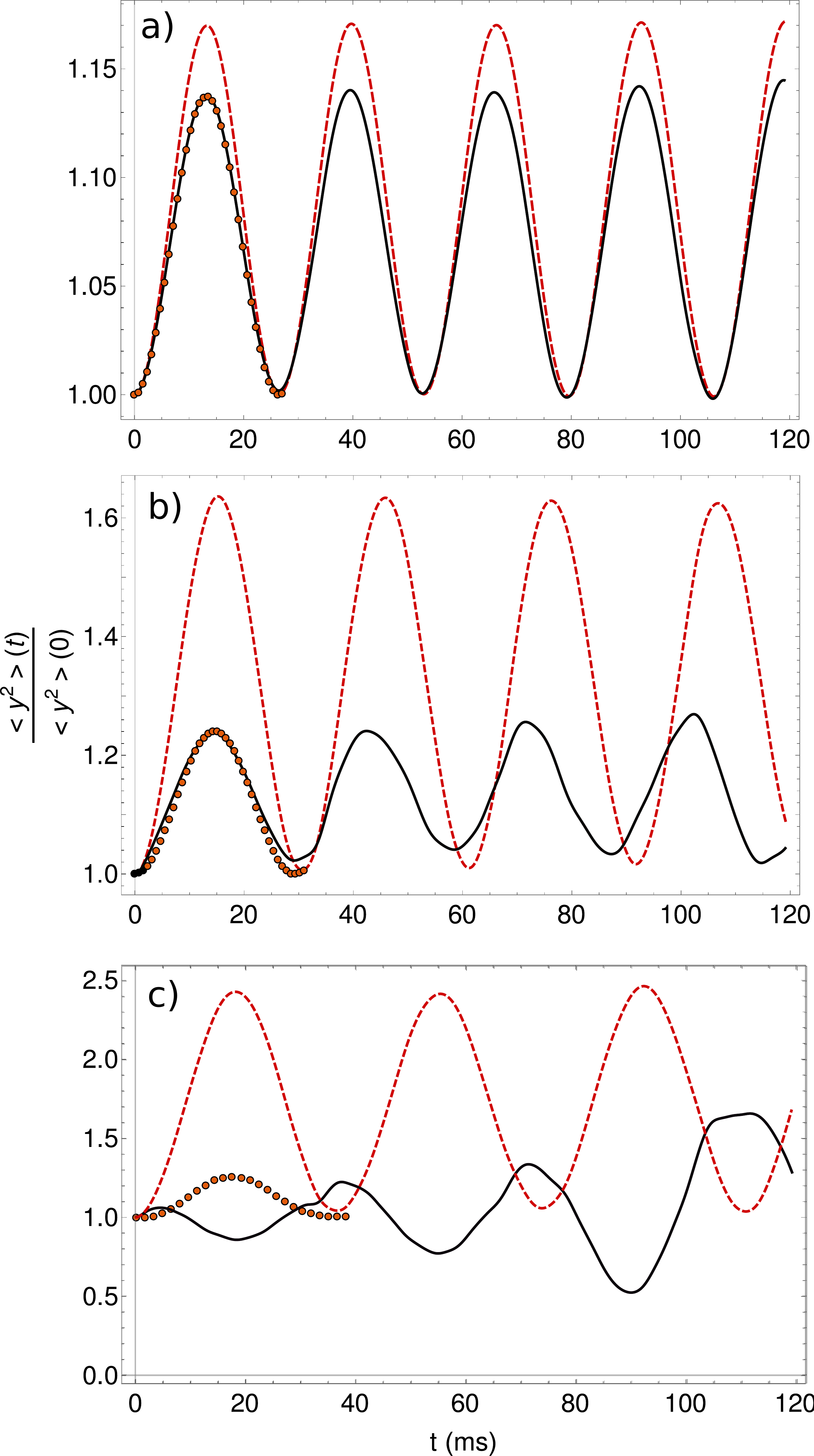}}
\caption{The square of the relative distance between the impurities as a function of time for repulsive impurity-boson interactions. We choose to normalize the distance to one at $t=0$. The (red) dashed curve in every panel shows $[\langle y_1^2 + y_2^2 \rangle_{1,\mathrm{MB}} - \langle {y_{\mathrm{CM}}}^2 \rangle_{\mathrm{eff}}](t)/[\langle y_1^2 + y_2^2 \rangle_{1,\mathrm{MB}} - \langle {y_{\mathrm{CM}}}^2 \rangle_{\mathrm{eff}}] (t=0)$  for two uncorrelated impurities~\cite{mistakidis2018}. The (black) solid curve shows the ML-MCTDHB results for two correlated impurities, $[\langle y_1^2 + y_2^2 \rangle_{\mathrm{MB}} - \langle {y_{\mathrm{CM}}}^2 \rangle_{\mathrm{eff}}](t)/[\langle y_1^2 + y_2^2 \rangle_{\mathrm{MB}} - \langle {y_{\mathrm{CM}}}^2 \rangle_{\mathrm{eff}}] (t=0)$ . The dots in panels {\bf a)} and {\bf b)} show the best fits to the effective Hamiltonian~(\ref{eq:eff_ham_rel}); the rightmost dot in every panel indicates the time interval used for the fitting. The fitted parameters are $g_{II}^i=-0.011 g_{BB}$ in panel {\bf a)}  and $g_{II}^i=-0.10 g_{BB}$ in panel {\bf b)}. The dots in panel {\bf c)} show the result of the effective model with $g_{II}^i=-0.25 g_{BB}$. All panels are for impurities that do not interact in free space, $g_{II}=0$. In panel {\bf a)} we use $g_{IB}(t>0)=0.1 g_{BB}$, in panel {\bf b)} $g_{IB}(t>0)=0.3 g_{BB}$, and in panel {\bf c)} $g_{IB}(t>0)=0.5g_{BB}$.   
}
\label{fig:fig2}
\end{figure}
%%%%%%%%%%%%%%%%%%%%%%%%%%%%%%%%%%%%%%%%%%%%%%%%%%%%%%%%%%%%%%%%%%%%%%%%%%%%%%%%%%%%%%%%%%%%%%%%%%%%%%%%%%%%%%%%%%%%%%%

For the sake of discussion, we consider only systems with $g_{II}=0$, which enjoy the most direct and clear manifestation of induced interactions. Indeed, in this case all observed correlations are induced. We determine $\langle y_1^2 + y_2^2 \rangle_{\mathrm{MB}}$ using the ML-MCTDHB method and compare it with $\langle y^2+y_{\mathrm{CM}}^2  \rangle_{\mathrm{eff}}$ from the effective Hamiltonian [Eq.~(\ref{eq:eff_ham_rel})]. To isolate the effect of the induced interactions, we exclude the contribution of the center-of-mass motion and compare the time evolution of $\langle y_1^2 + y_2^2 \rangle_{\mathrm{MB}} - \langle {y_\mathrm{CM}}^2 \rangle_{\mathrm{eff}}$ and $\langle y^2 \rangle_{\mathrm{eff}}$. The former quantity describes the dynamics of the relative distance between the two impurities only approximately, because the center-of-mass and the relative motions of the impurities are coupled in the many-body calculations via interactions with the bath. The parameter $g_{II}^i$ for the effective potential is obtained by fitting to the corresponding ML-MCTDHB data, and then compared to Eq.~(\ref{eq:est_gIIi}). To minimize the role of the effects arising due to the in-homogeneity of the Bose gas, we only fit to the data that describe the first oscillation. For consistency, the effective Hamiltonian is used throughout this work to describe the time evolution only during the first oscillation period of $\langle y^2\rangle_{\mathrm{eff}}$ (e.g., $\simeq 28 ms$ for Fig.~\ref{fig:fig2}{\bf a)}). The discussion does not change qualitatively if we use for the fitting the first two or even three oscillation periods. 

\begin{figure}
  	\includegraphics[width=0.47\textwidth]{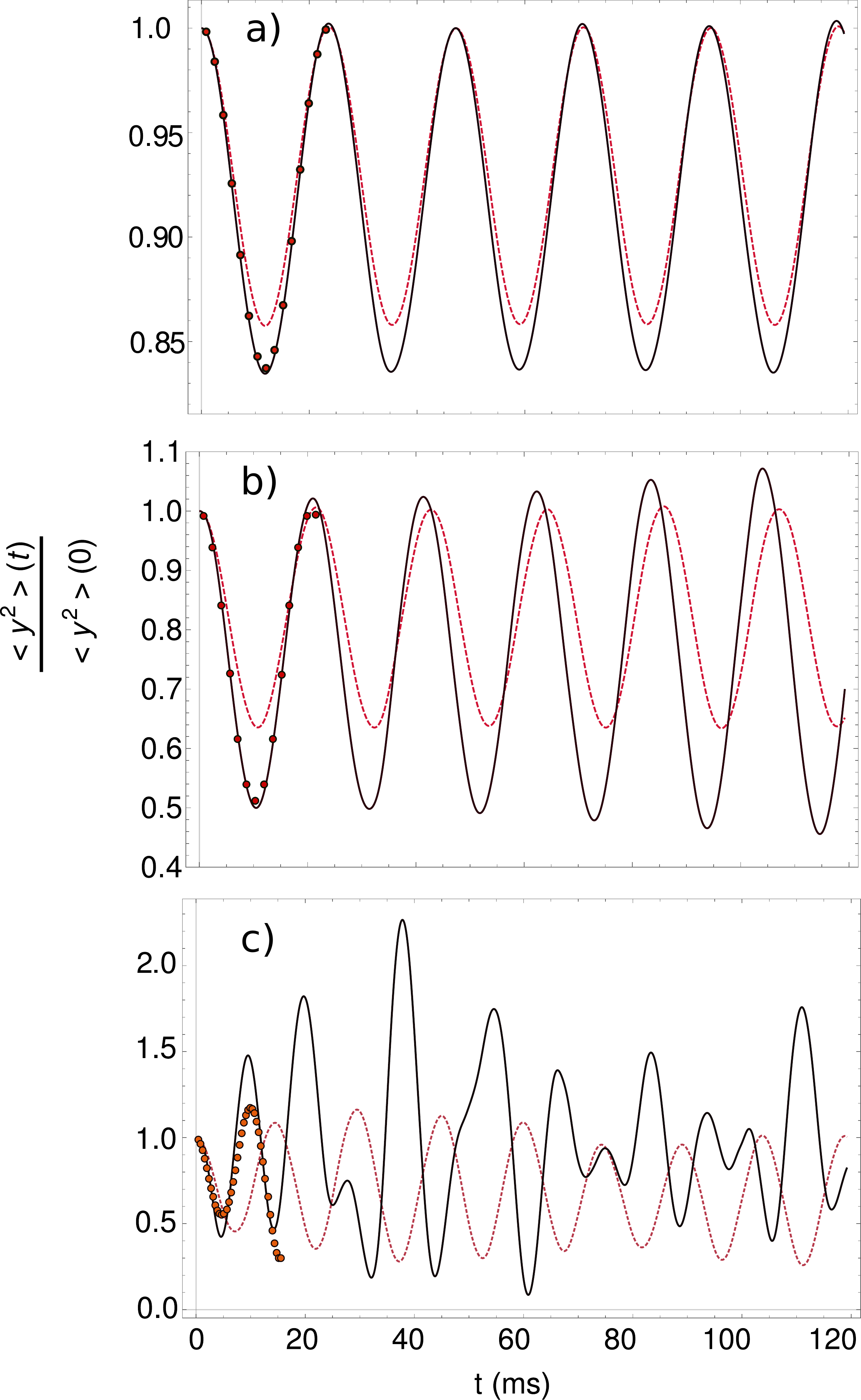}
  	\caption{ The square of the relative distance between the impurities as a function of time for attractive impurity-boson interactions. We choose to normalize the distance to one at $t=0$. The (red) dashed curve in every panel shows $[\langle y_1^2 + y_2^2 \rangle_{1,\mathrm{MB}} - \langle {y_{\mathrm{CM}}}^2 \rangle_{\mathrm{eff}}](t)/[\langle y_1^2 + y_2^2 \rangle_{1,\mathrm{MB}} - \langle {y_{\mathrm{CM}}}^2 \rangle_{\mathrm{eff}}] (t=0)$  for two uncorrelated impurities~\cite{mistakidis2018}. The (black) solid curve shows the ML-MCTDHB results for two correlated impurities, $[\langle y_1^2 + y_2^2 \rangle_{\mathrm{MB}} - \langle {y_{\mathrm{CM}}}^2 \rangle_{\mathrm{eff}}](t)/[\langle y_1^2 + y_2^2 \rangle_{\mathrm{MB}} - \langle {y_{\mathrm{CM}}}^2 \rangle_{\mathrm{eff}}] (t=0)$.  The dots in the panels show the results of the effective Hamiltonian~(\ref{eq:eff_ham_rel}) with $g_{II}^i=-0.011 g_{BB}$  (panel {\bf a)}),  $g_{II}^i=-0.10 g_{BB}$  (panel {\bf b)}), and $g_{II}^i=-0.74 g_{BB}$  (panel {\bf c)}). The impurities do not interact in free space, $g_{II}=0$. In panel {\bf a)} we use $g_{IB}(t>0)=-0.1 g_{BB}$; in panel {\bf b)} $g_{IB}(t>0)=-0.3 g_{BB}$; in panel {\bf c)} $g_{IB}(t>0)=- g_{BB}$.  }
  	\label{fig:negat_quenches} 
\end{figure}

Our results are presented in Fig.~\ref{fig:fig2}. 
First of all, we observe that the values of $g_{II}^i$ used in Fig.~\ref{fig:fig2}{\bf a)} and {\bf b)} are consistent with the estimate of Eq.~(\ref{eq:est_gIIi}). For the case $g_{IB}=0.1g_{BB}$, Eq.~(\ref{eq:est_gIIi}) suggests that $g_{II}^i\simeq -0.01g_{BB}$; for $g_{IB}=0.3g_{BB}$ it implies $g_{II}^i\simeq -0.09g_{BB}$. These estimates are in a good agreement with the values obtained by fitting to the ML-MCTDHB data, namely: $g_{II}^i=-0.011g_{BB}$ and $g_{II}^i=-0.10g_{BB}$. We observe that even though the parameter $g_{II}^i$ is small it significantly suppresses the amplitude of the oscillations, compare the dashed and solid curves in Fig.~\ref{fig:fig2}. Therefore, the induced correlations could be studied in an experiment  by monitoring the quench dynamics. The challenge here is to create the initial state, $\psi(t=0)$, at sufficiently low temperatures (see also Sec.~\ref{sec:summary}). We also explore the change of the boson-impurity interaction strength to negative values of $g_{IB}$, i.e., $g_{IB}(t>0)<0$. According to Eq.~(\ref{eq:est_gIIi}), the strength of induced correlations is independent of the sign of $g_{IB}$, if $g_{BB}$ and $|g_{IB}|$ are small. The ML-MCTDHB simulations agree with this conclusion, see Fig.~\ref{fig:negat_quenches}. The dots in Fig.~\ref{fig:negat_quenches} show not a fit, but a prediction based on $g_{II}^i$ obtained for $g_{IB}>0$, e.g, we use in the effective model $g_{II}^i=-0.1g_{BB}$ fitted above for $g_{IB}(t>0)=0.3g_{BB}$ to obtain the dynamics following the change to $g_{IB}(t>0)=-0.3g_{BB}$.  The parameters $\overline m_{\mathrm{eff}}$ and $\overline k$ for Eqs.~(\ref{eq:h1_trap})~and~(\ref{eq:eff_ham_rel}) are obtained  as in Ref.~\cite{mistakidis2018}, by fitting to the ML-MCTDHB results for a single impurity; these parameters are $\overline m_{\mathrm{eff}}=1.02 m, \overline k =1.14 k$ for $g_{IB}(t>0)=-0.1 g_{BB}$, and $\overline m_{\mathrm{eff}} =1.075 m, \overline k =1.46 k$ for $g_{IB}(t>0)=-0.3 g_{BB}$. All in all, the results shown in Figs.~\ref{fig:negat_quenches}{\bf a)} and {\bf b)} confirm our expectations based on Fig.~\ref{fig:fig2}~: the relative distance for two weakly-correlated impurities is smaller than the distance between two uncorrelated impurities, which we attribute to the attractive interaction mediated by the Bose gas.  

 For larger values of $|g_{IB}|$ the quench dynamics for $g_{IB}>0$ cannot be compared with that for $g_{IB}<0$. For example, for $g_{IB}=g_{BB}$, the impurities are pushed to the edge of the trap, whereas for $g_{IB}=-g_{BB}$, the impurities move in the vicinity of the centre of the trap, see Fig.~\ref{fig:negat_quenches}~{\bf c)}. For comparison, we present in Fig.~\ref{fig:negat_quenches}~{\bf c)} also the results of the effective model with $\overline m_{\mathrm{eff}}=0.83 m, \overline k =2.56 k$ and $g_{II}^i=-0.74g_{BB}$. As before, the parameters $\overline m_{\mathrm{eff}}$ and $\overline k$ are obtained from the dynamics of a single impurity in a Bose gas. The parameter $g_{II}^i$ is now obtained by fitting to the ML-MCTDHB results for two impurities.

We remark that already for $g_{IB}=0.5 g_{BB}$ the data cannot be accurately fitted with the proposed effective model. For the sake of discussion, we show the result of the effective model with $g_{II}^i=-0.25g_{BB}$, which is expected from Eq.~(\ref{eq:est_gIIi}), in Fig.~\ref{fig:fig2}{\bf c)}. Our failure to fit the data is not surprising: (i) The impurity starts to probe a large part of the Bose cloud [see the amplitude of oscillation of the non-correlated impurities in Fig.~\ref{fig:fig2}{\bf c)}], which means that our assumption that $V$ depends only on $y_1-y_2$ must be modified; (ii) The zero-range approximation to the effective potential is also not correct, impurities start to resolve the shape of the induced potential. Some of these effects might be included by using a more elaborate form of the effective potential, as, e.g., in Refs.~\cite{dehkharghani2018, chen2018}. This discussion is beyond the scope of the present paper, therefore, we refrain from discussing cases with $g_{IB}>0.3g_{BB}$ further.

%%%%%%%%%%%%%%%%%%%%%%%%%%%%%%%%%%%%%%%%

%%%%%%%%%%%%%%%%%%%%%%%%%%%%%%%%%%%%%%%%

\subsection{The entropy}

Induced impurity-impurity correlations can be studied using the
one-body reduced density matrix 
\begin{align}
\rho^{(I)}_{MB}&(y_1,y_1';t)=\int\mathrm{d}y_2\mathrm{d}x_1...\mathrm{d}x_N&\nonumber\\
&\Psi_{MB}(y_1,y_2,x_1,...,x_N; t)\Psi_{MB}^*(y_1',y_2,x_1,...,x_N; t).
\end{align} 
Let us denote the eigenvalues of $\rho^{(I)}_{{MB}}$ as $n_i(t)$. We arrange them in ascending order, such that $n_1\geq n_2\geq n_3 \geq...$;  $n_i(t)$ are often referred to as natural occupation numbers. They can be seen as probabilities for finding an impurity in a given state, since $\sum_i n_i=1$.
It is clear that the impurities are correlated if two or more $n_i$ are non-zero -- otherwise, the wave function, $\Psi_{MB}$, is just a product state.
To quantify the strength of correlations, we introduce the entropy  
\begin{align}
S(t)=-\sum\limits_{i=1} n_{i}(t)\ln[n_i(t)].\label{eq:entropy_impurity} 
\end{align} 
If $n_1(t)=1$ and $n_{j\neq 1}(t)=0$ then $S(t)=0$, signaling the absence of impurity-impurity correlations. Other possibilities lead to non-zero values of $S(t)$.  
Since the two bosonic impurities are non-interacting in free space, any positive value of entropy can be used as a witness of induced impurity-impurity correlations  
mediated by the bosonic gas.

\begin{figure}
  	\includegraphics[width=0.47\textwidth]{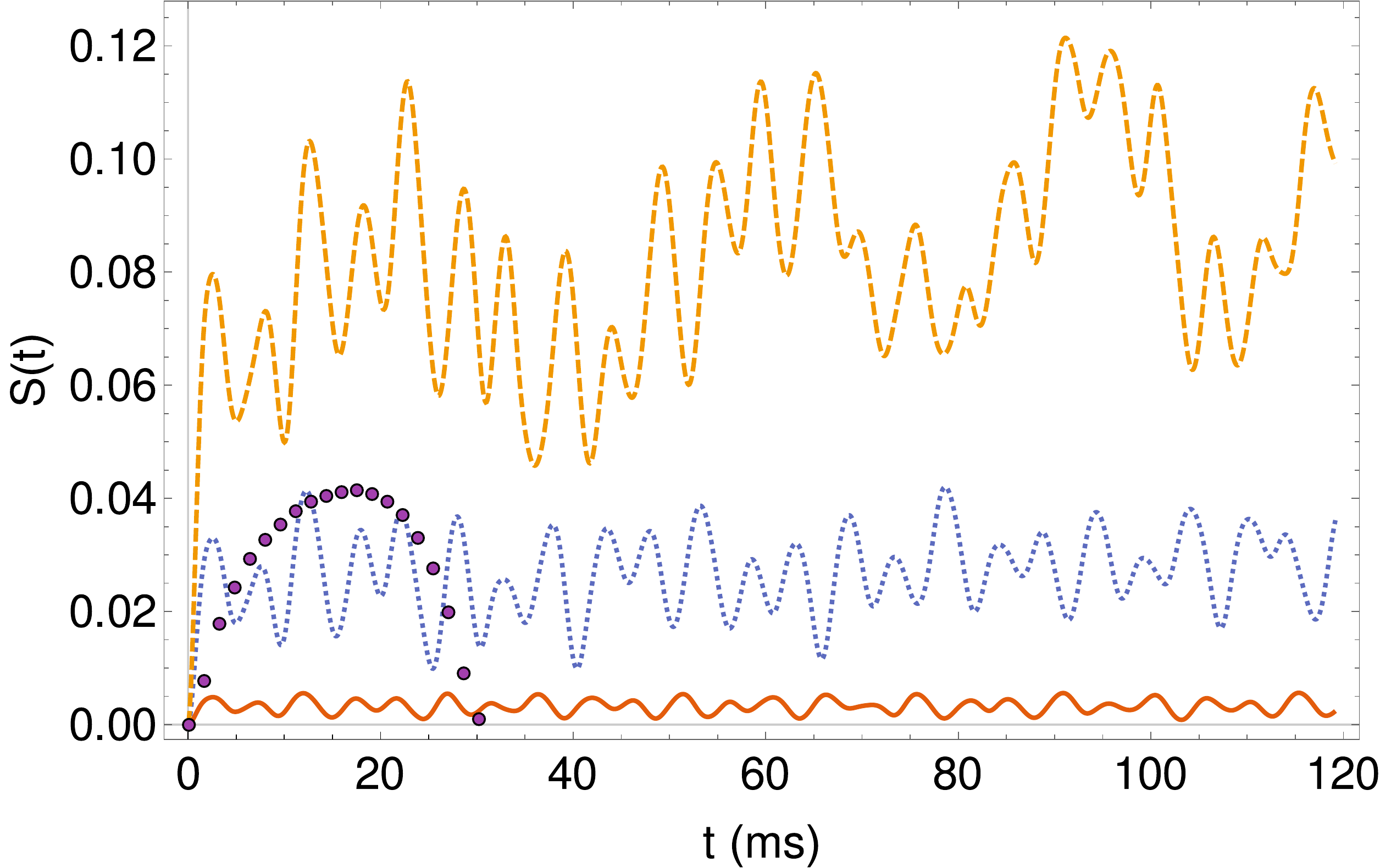}
  	\caption{The entropy, $S(t)$, as a function of time for the quench dynamics following the change: $g_{IB}(t=0)=0$ to $g_{IB}(t>0)\neq 0$. The lower (solid) curve describes $g_{IB}=0.1g_{BB}$, the middle (dotted) curve is for $g_{IB}=0.3g_{BB}$, and the upper (dashed) curve is for $g_{IB}=0.5 g_{BB}$.
  	The dots show the result of the effective model with $g_{II}^i=-0.10g_{BB}$. This value of $g_{II}^i$ was obtained from the ML-MCTDHB data in Fig.~\ref{fig:fig2}{\bf b)}, see the text for details. }
  	\label{fig:entropy} 
\end{figure}

We first calculate the time evolution of $S(t)$ using the ML-MCTDHB method for different values of $g_{IB}$ for $t>0$, see Fig.~\ref{fig:entropy}.
Initially, the impurities do not interact, therefore, $S(t=0)=0$.
For $t>0$ we observe that $S\neq 0$, which indicates the presence of impurity-impurity correlations.  
For $g_{IB}(t>0)=0.1$, the entropy is close to zero, meaning that the mean-field treatment can be used to decently describe the dynamics for such weak interactions. For larger values of $g_{IB}(t>0)$, beyond-mean-field corrections are important and must be taken into account (cf. Appendix~\ref{methodology}, where mean-field calculations of the size of the impurity cloud are presented).
At the early stages of the dynamics, $S(t)$ exhibits a sudden linear increase, while, for later time-instants, it shows an oscillatory behavior 
possessing a multitude of frequencies. 
The initial increase of $S(t)$ is more pronounced for larger postquench interspecies interaction strengths, e.g., $S_I(t=3.3)=0.032$ for $g_{IB}=0.3 g_{BB}$,
and $S(t=3.3)=0.079$ for $g_{IB}=0.5 g_{BB}$, suggesting stronger impurity-impurity correlations for larger values of $g_{IB}$. 
Also the average degree of impurity-impurity correlations quantified by $\bar{S}(T)=(1/T)\int_0^T dt S(t)$, is 
larger for an increasing postquench $g_{IB}$, for instance, $\bar{S}(120ms)\approx 0.027$ at $g_{IB}=0.3 g_{BB}$, and $\bar{S}(120ms)\approx 0.082$ at $g_{IB}=0.5 g_{BB}$. 
We conclude that a larger postquench $g_{IB}$ implies a larger degree of impurity-impurity correlations~(cf.~\cite{mistakidis2019,BF_dynamics,expansion}). 

%%%%%%%%%%%%%%%%%%%%%%%%%%%%%%%%%%%%%%%%

%%%%%%%%%%%%%%%%%%%%%%%%%%%%%%%%%%%%%%%%

Next we calculate the entropy using the effective model [Eq.~(\ref{eq:eff_ham_rel})]. To this end, we diagonalize the one-body density matrix,
\begin{equation}
\rho_{\mathrm{eff}}^{(I)}(y_1,y_1';t)=\int \mathrm{d}y_2 \Psi_{\mathrm{eff}}(y_1,y_2;t)\Psi^*_{\mathrm{eff}}(y_1',y_2;t),
\end{equation}
whose eigenvalues are then used in Eq.~(\ref{eq:entropy_impurity}) to determine the entropy, see Appendix~\ref{sec:app_a} for details. 
We show this entropy for $g_{II}^i=-0.10g_{BB}$ [$g_{IB}(t>0)=0.3g_{BB}$] in Fig.~\ref{fig:entropy} for up to $t\simeq 30$ms, which corresponds to one oscillation period (cf.~Fig.~\ref{fig:fig2}{\bf b)}).
The spectrum of the effective model is close to that of a harmonic oscillator. It leads to an almost complete restoration of the initial wave packet, and, hence, $S(t)$ approaches a minimal value after one oscillation period. In contrast, the ML-MCTDHB results suggest a fast approach to a thermal-like state, where the entropy oscillates around its average value.  However, the prediction of the ML-MCTDHB and the effective model agree on the average values of the entropies.  The ML-MCTDHB method yields
$\overline S(30ms)\simeq 0.025$ for $g_{IB}=0.3 g_{BB}$, which should be compared with $\overline S(30ms)\simeq 0.029$ of the effective model. We conclude that the effective model is useful to calculate average values of the entropy. We checked that this conclusion holds also for the dynamics initiated by a change to attractive boson-impurity interactions, i.e., if $g_{IB}(t>0)<0$.

To give insight into the difference between the entropy calculated using the many-body ML-MCTDHB approach and the effective model, we note that the dynamics in the effective model is driven by the harmonic oscillator whose equidistant spectrum gives a single oscillation frequency. In reality, there is a multitude of processes associated with the impurity-bath interactions, which affect the equidistant spectrum of the harmonic oscillator. The ML-MCTDHB calculations (unlike the effective model) capture these processes, which lead to the observed thermal-like state. In the future, it will be interesting to study possible extensions of the effective model that can lead to the observed approach to a thermal-like state.

\subsection{The two-body correlation function}
\label{subsec:two_body_correl}

To reveal impurity-impurity correlations in a spatially resolved manner, we study the normalized 
two-body correlation function
\begin{align}
g^{(II)}_{MB}(y_1,y_2;t)=\frac{\rho_{MB}^{(II)}(y_1,y_2;t)}{\rho^{(I)}_{MB}(y_1,y_1;t)\rho_{MB}^{(I)}(y_2,y_2;t)}, \label{two_body_correlation}
\end{align}
where $\rho^{(II)}_{MB}(y_1,y_2;t)$ is the two-body reduced density matrix 
\begin{align}
\rho^{(II)}_{MB}(y_1,y_2;t)=\int\mathrm{d}x_1...\mathrm{d}x_N|\Psi_{MB}(y_1,y_2,x_1,...,x_N; t)|^2.
\end{align} 
The function $\rho^{(II)}_{MB}(y_1,y_2;t)$ represents the probability of measuring at time $t$ one impurity at $y_1$ and another at $y_2$.  The two impurities correlate if $g^{(II)}_{MB}(y_1,y_2;t)\neq 1$; $g^{(II)}_{MB}(y_1,y_2)>1$ ($g^{(II)}_{MB}(y_1,y_2)<1$) implies an increased (decreased) probability to observe two particles at $y_1$ and $y_2$ in comparison to two uncorrelated impurities. If $g^{(II)}_{MB}(y_1,y_2;t)=1$ the impurities are  uncorrelated.

%%%%%%%%%%%%%%%%%%%%%%%%%%%%%%%%%%%%%%%
\begin{figure}
  	\includegraphics[width=0.5\textwidth]{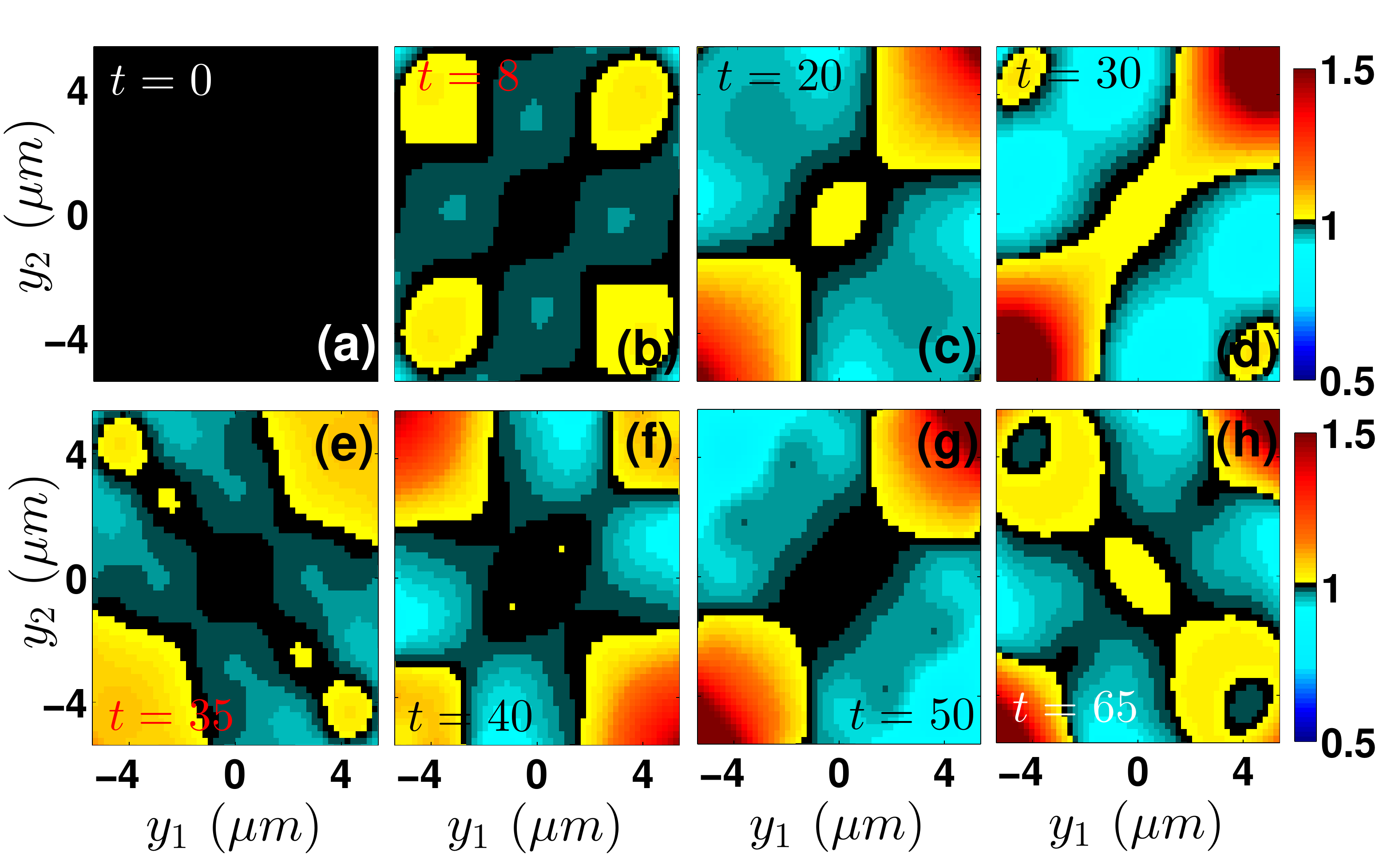}
  	\caption{The two-body correlation function, $g_{MB}^{(II)}$, of the two bosonic impurities at different time instants in $ms$ (see legends) during the quench dynamics that 
  	follows the change of the boson-impurity interaction strength: $g_{IB}(t=0)=0$ to $g_{IB}(t>0)=0.3 g_{BB}$. }
  	\label{fig:two_body_coherence} 
\end{figure} 
%%%%%%%%%%%%%%%%%%%%%%%%%%%%%%%%%%%%%%%

Figure~\ref{fig:two_body_coherence} shows $g^{(II)}_{MB}(y_1,y_2;t)$ for the change of
the interaction strength from $g_{IB}=0$ to $g_{IB}=0.3g_{BB}$. 
Initially, at $t=0$, the two non-interacting bosonic impurities are uncorrelated which leads to $g^{(II)}_{MB}(y_1,y_2;t=0)=1$ [Fig. \ref{fig:two_body_coherence} {\bf (a)}]. 
Two-body correlations between the impurities begin to develop at $t>0$ [Fig. \ref{fig:two_body_coherence} {\bf (b)}] and become more 
pronounced as time evolves [Figs.~\ref{fig:two_body_coherence}~{\bf (c)-(h)}]. 
Recall that the two bosonic impurities perform a breathing motion in the course of the time evolution [cf. the oscillations in Fig.~\ref{fig:fig2}{\bf b)}]. 
This breathing dynamics leads to the expansion and contraction of $\rho^{(I)}(y_1,y_1;t)$, and to two distinct correlation patterns present in $g^{(II)}_{MB}$. We exemplify them in Figs.~\ref{fig:two_body_coherence}~{\bf (c)}~and~{\bf (f)}. In Fig.~\ref{fig:two_body_coherence}~{\bf (c)}, the two impurities attract each other and move together to the edge of the cloud, since $g^{(II)}_{MB}(y_1,y_2=y_1;t=20ms)\geq 1$. In Fig.~\ref{fig:two_body_coherence}{\bf (f)}, the impurities also move towards the edge of the trap. However, they move not only as a pair but also separately, because  both $g^{(II)}_{MB}(y_1,y_2=y_1;t=40ms)$ and $g^{(II)}_{MB}(y_1,y_2=-y_1;t=40ms)$ are greater than (or equal to) one. 

\begin{figure}
  	\includegraphics[width=0.5\textwidth]{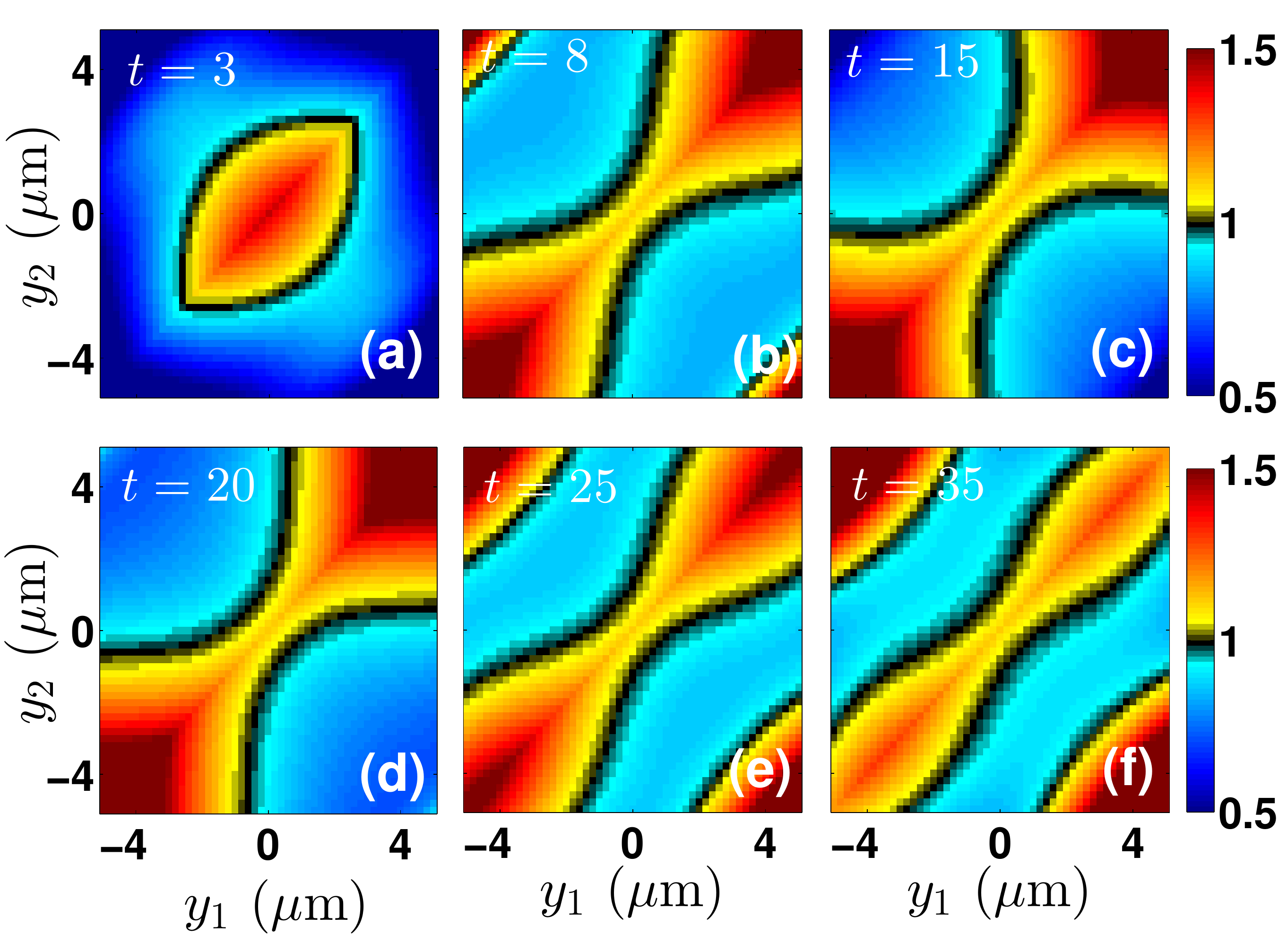}
  	\caption{The two-body correlation function, $g_{\mathrm{eff}}^{(II)}$, of the two bosonic impurities at different time instants in $ms$ (see legends) during the quench dynamics upon the change of the induced impurity-impurity interaction strength in the effective model that corresponds to $g_{IB}(t>0)=0.3 g_{BB}$. The figure should be compared to Fig.~\ref{fig:two_body_coherence}.}
  	\label{fig:two_body_coherence_eff} 
\end{figure}

To better understand impurity-impurity correlations depicted in Fig.~\ref{fig:two_body_coherence}, we use the effective model to calculate the two-body correlation function 
\begin{equation}
g_{\mathrm{eff}}^{(II)} = \frac{|\Psi_{\mathrm{eff}}(y_1,y_2;t)|^2}{\int |\Psi_{\mathrm{eff}}(y_1,y_2;t)|^2\mathrm{d}y_1 \int |\Psi_{\mathrm{eff}}(y_1,y_2;t)|^2\mathrm{d}y_2}.
\end{equation}
The time evolution of $g_{\mathrm{eff}}^{(II)}$ is depicted in Fig.~\ref{fig:two_body_coherence_eff} for $g_{IB}(t>0)=0.3 g_{BB}$. We observe a qualitative agreement between Figs.~\ref{fig:two_body_coherence} and~\ref{fig:two_body_coherence_eff}, which allows us to use the effective model to interpret the behavior of the ML-MCTDHB data. For instance, similarly to $g_{MB}^{(II)}$ [Fig.~\ref{fig:two_body_coherence}{\bf (b)}, {\bf (e)}], the function $g_{\mathrm{eff}}^{(II)}$ [Fig.~\ref{fig:two_body_coherence_eff}{\bf (b)}, {\bf (e)}] features two distinct patterns that are characterized by $g_{\mathrm{eff}}^{(II)}>1$ along the diagonal and $g_{\mathrm{eff}}^{(II)}>1$ across the antidiagonal. According to the effective model, these patterns represent the motion of a bound pair and the backward scattering of two dressed impurity atoms, respectively.

%%%%%%%%%%%%%%%%%%%%%%%%%%%%%%%%%%%%%%%
%%%%%%%%%%%%%%%%%%%%%%%%%%%%%%%%%%%%%%%

\section{Summary and Outlook}
\label{sec:summary}

We explore the time evolution of a Bose gas with two impurities initiated by a sudden change of the boson-impurity interaction strength. Our focus is on the dynamics of the impurities, which we analyze using the many-body correlated ML-MCTDHB method~\cite{Mistakidis_cor, Katsimiga_DBs} and an effective model presented in Eq.~(\ref{eq:ham_eff_overline}). We observe that the Bose gas induces impurity-impurity correlations. The strength of the induced interactions can be estimated from the attractive Yukawa-type potential between two heavy impurities in a homogeneous Bose gas~\cite{recati2005, bruderer2007,dehkharghani2018,reichert2018}. This potential captures well the dynamics of the size of the impurity cloud. It also explains qualitatively the time-averaged value of the entropy and the correlation patterns that appear in the two-body correlation function. We observe a pronounced effect of the induced correlations on the dynamics. This means that the quench dynamics may become a testground for studies on induced interactions provided that the initial state, the ground state of the system, can be prepared.

It will be interesting in the future to explore other types of induced interactions. For example, the long-range interactions ($1/r^3$) due to quantum fluctiations are too weak in our system, but can become important if two impurities are confined by traps whose origins are well-separated. Another exciting direction is systems with repulsive induced interactions. Repulsive interactions can appear if two impurities are distinguishable with one impurity repelling the Bose gas, and the second impurity attracting it. Then, according to Eq.~(\ref{eq:density_with_imp}), the impurity-impurity interaction is repulsive; its strength is given by $g_{II}^i=-g_{IB}^1g_{IB}^2/g_{BB}$, here $g_{IB}^l$ characterizes the interaction between a boson and the $l$th impurity. 

Future studies are needed to understand the effect of a finite temperature, $T$, on the discussed dynamics.  The energy scale associated with the induced interaction is small (see the estimate below) in comparison to typical energy scales in current cold-atom experiments. Therefore, it is natural to expect that the induced correlations are important only at very low temperatures (cf.~Ref.~\cite{reichert2019, pavlov2019}). To estimate the relevant temperature scales, we analyze a dimensionless parameter constructed from quantities that enter~Eq.~(\ref{eq:eff_ham_rel}): $s=|m(g_{II}^{(i)}+g_{II})\lambda_{typ}/(\hbar^2)|$, where $\lambda_{typ}$ is a typical length scale. The parameter $s$ is the only dimensionless parameter for a homogeneous system, it also determines the relative strength of the induced interactions in trapped systems. If $s$ is large then the induced interactions are important; in the opposite limit, $s\to 0$, the impurities correlate weakly. If we use a thermal wave length as a typical length scale, i.e., $\lambda_{typ}\simeq \hbar/\sqrt{2m k_B T}$ ($k_B$ is the Boltzmann constant), then the parameter $s$ is close to unity for $T\simeq n$K, assuming that $g_{II}+g_{II}^{i}$ are in the range $-(10^{-38} \div 10^{-37}) Jm$, and $m=m(^{87}{\mathrm{Rb}})$. For higher temperatures the induced interactions cannot affect the dynamics, since  $s\to 0$. Future studies are needed to understand the transition from small to large values of $s$.  

In this work, we focus on the case with $g_{II}=0$. This is an ideal scenario, which might be difficult to realize experimentally. In the future, it will be interesting to identify the most promising systems in which induced interactions can be observed even if $g_{II}\neq 0$. 
Note that the limit $1/g_{II}\to 0$, which corresponds to fermionic impurities, has already been studied~\cite{BF_dynamics,dehkharghani2018, pasek2019}. This limit features much weaker effect of the induced interactions on the properties of the system.

In the present paper we focus on the dynamics of the impurities. However, the ML-MCTDHB method also gives access to the dynamics of the Bose gas, which, for weak interactions considered here, performs a small-amplitude breathing motion at $t>0$.
In general, we expect a number of interesting phenomena associated with the Bose gas. For example, it is known that switching off boson-impurity interactions can lead to shock waves and solitons in the Bose gas~\cite{akram2016}. In light of the importance of beyond-mean-field effects in our study [see Appendix~\ref{methodology}], it will be interesting to study these non-linear objects for moderate interactions using the ML-MCTDHB method~\cite{Mistakidis_cor, Katsimiga_DBs}.

\vspace*{1em}

\begin{acknowledgments}
We thank Nikolaj Zinner for many inspiring discussions regarding a Bose gas with impurity atoms.  
 This work has received funding from the Deutsche Forschungsgemeinschaft (DFG) 
in the framework of the SFB 925 ``Light induced dynamics and control of correlated quantum systems'' (S.~I.~M. and P.~S.);
 DFG Project No.413495248 [VO 2437/1-1] and European Union's Horizon 2020 research and innovation programme under the Marie Sk\l{}odowska-Curie Grant Agreement No. 754411 (A. G. V.).   S.~I.~M. gratefully acknowledges financial support in the framework of the Lenz-Ising Award 
of the Department of Physics of the University of Hamburg.

\vspace*{1em}

S.~I. M. and A.~G. V. contributed equally to this work.
\end{acknowledgments}

\vspace*{1em}

\appendix
\widetext

\section{A brief description of the ML-MCTDHB method} \label{methodology}

To investigate the stationary properties of the Hamiltonian in Eq.~(\ref{eq:ham}), and the quench dynamics we address numerically the underlying many-body Schr{\"o}dinger equation. We employ the ML-MCTDHB method~\cite{kronke2013, cao2013,mistakidis2017}, which is a variational approach for solving the time-dependent many-body Schr{\"o}dinger equation for atomic mixtures composed either 
of bosonic \cite{Mistakidis_cor,mistakidis2018,Katsimiga_DBs,mistakidis2019} or fermionic 
\cite{Erdmann_tunel,Erdmann_phase_sep,Koutentakis_FF,expansion,BF_dynamics,Mistakidis_fermi_pol} species. The method relies on the expansion of the many-body wave function in terms of a time-dependent and variationally optimized 
basis enabling us to take into account both the inter- and the intraspecies correlations of two-component systems. In this Appendix, we briefly review the method for the convenience of the reader.

The many-body wave function is first expressed as a superposition of $D$ different species functions for each of the species, i.e.  
$\Psi^{B}_k (\vec x;t)$ and $\Psi^{I}_k (\vec y;t)$.  
Here, $\vec x=\left( x_1, \dots, x_{N} \right)$ and $\vec y=\left( y_1, y_{2} \right)$ are the spatial coordinates of the $N$ bosons and the two impurities, respectively. 
Consequently, the many-body wavefunction $\Psi_{MB}(t)$ includes interspecies correlations and has the form of a truncated Schmidt 
decomposition~\cite{Horodecki} with rank $D$, namely  
\begin{equation}
\Psi_{MB}(t) = \sum_{k=1}^D \sqrt{ \lambda_k(t) } \Psi^B_k (t) \Psi^I_k (t).    
\label{Eq:WF}
\end{equation}
The Schmidt coefficients $\lambda_k(t)$ are known as the natural species populations of the $k$-th species 
function~\cite{Mistakidis_cor,mistakidis2019,mistakidis2017}. 
The system is said to be entangled~\cite{Roncaglia} or interspecies correlated when at least two 
distinct $\lambda_k(t)$ possess a nonzero value. 
In this case $\Psi_{MB}$ cannot be expressed as a direct product of two states. 
Note also that $\lambda_k(t)$ are the eigenvalues of the species reduced density matrices,
e.g., $\rho^{B} (\vec{x}, \vec{x}';t)=\int \mathrm{d} y_1 \mathrm{d} y_2 \Psi^*_{MB}(\vec{x}, 
\vec{y};t) \Psi_{MB}(\vec{x}',\vec{y};t)$.  
The function $\Psi^{B}_k (\vec x;t)$ [resp.~$\Psi^{I}_k (\vec y;t)$] is further expanded in the basis made of $d_{B}$ [resp.~$d_{I}$] distinct time-dependent single-particle functions (SPFs) namely $\{\varphi_1^{B},\dots,\varphi_{d_{B}}^{B}\}$ [resp.~$\{\varphi_1^{I},\dots,\varphi_{d_{I}}^{I}\}$]. In other words, we write the function $\Psi^{B}_k (\vec x;t)$ [resp.~$\Psi^{I}_k (\vec y;t)$] as a linear combination of time-dependent number-states, $\vec{n}^{B} (t)$ [resp.~$\vec{n}^{I} (t)$], 
with time-dependent expansion coefficients $C^{B}_{k;\vec{n}}(t)$ [$C^{I}_{k;\vec{n}}(t)$]:
\begin{equation}
     \Psi_k^{B} (t) =\sum_{\vec{n}} C^{B}_{k;\vec{n}}(t) \vec{n}^{B} (t).  
    \label{Eq:SPF}
\end{equation} 
The number state $\vec{n}^{B} (t)$ is a time-dependent version of a basis state in the second quantization formalism. Namely, $\vec{n}^{B} (t)$ is a fully symmetric function constructed upon $d_{B}$ time-dependent variationally optimized SPFs, i.e., $\varphi_l^{B} (t)$, $l=1,2,\dots,d_{B}$ with occupation numbers $\vec{n}=(n_1,\dots,n_{d_{B}})$. 
A similar expansion holds for $\Psi^{I}_k (t)$ upon the change: $B\rightleftharpoons I$, $\vec{x}\rightleftharpoons\vec{y}$ and $d_{B}\rightleftharpoons d_{I}$. 
Additionally, the SPFs are expanded on a time-independent primitive basis $\lbrace \left| q \right\rangle \rbrace$ being in our case an $\mathcal{M}$ dimensional discrete 
variable representation (see also below).
Note that the eigenfunctions of the one-body reduced density matrix for bosons and impurities, i.e.,  
$\rho_\sigma^{(1)}(z,z^\prime;t)=\langle\Psi_{MB}(t)|\hat{\Psi}^{\sigma \dagger}(z)\hat{\Psi}^\sigma(z^\prime)|\Psi_{MB}(t)\rangle$ ($\sigma=\{B,I\}, z=\{x,y\}$), where $\hat{\Psi}^{\sigma}(z)$ is the bosonic field operator, are the so-called natural orbitals $\phi^{\sigma}_i(z;t)$. 
The natural orbitals are related with the SPFs via a unitary transformation that diagonalizes 
$\rho_\sigma^{(1)}(z,z^\prime;t)$ when it is expressed in the basis of SPFs. For a more elaborate discussion, see~Refs.~\cite{kronke2013, cao2013,mistakidis2017}. 
The eigenvalues of $\rho_\sigma^{(1)}(z,z^\prime;t)$ are the natural populations $n^{\sigma}_i(t)$. They provide a theoretical measure for the degree of intraspecies correlations. 
When there are more than one macroscopically occupied states, the $\sigma$-subsystem is called intraspecies correlated. Otherwise it is fully coherent. 

%%%%%%%%%%%%%%%%%%%%%%%%%%%%%%%%%%%%%%%
\begin{figure}
  	\includegraphics[width=\textwidth]{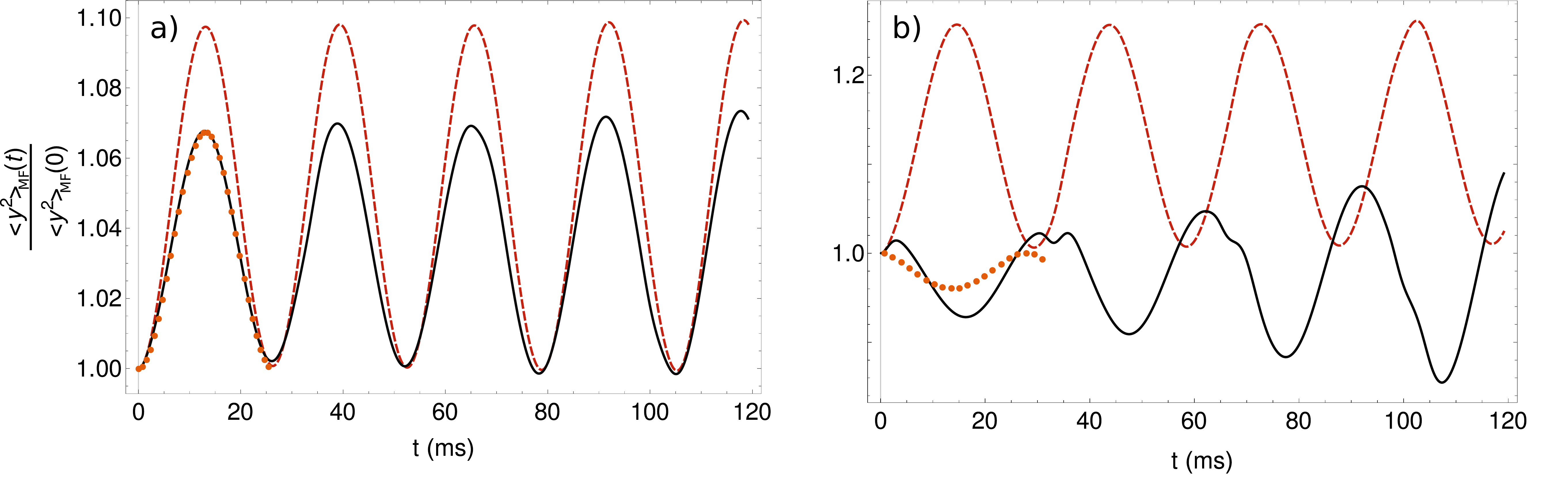}
  	\caption{The function $\langle y^2\rangle_{MF}(t)/\langle y^2\rangle_{MF}(0)$ as a function of time, $t$. The (red) dashed curve shows the mean-field result for impurities placed in two separate Bose gases~\cite{mistakidis2018}. The (black) solid curve shows the mean-field result for two impurities immersed together in a Bose gas. The dots show the result obtained from the effective Hamiltonian [see Eq.~(\ref{eq:eff_ham_rel})]. Panel {\bf a)} illustrates the dynamics with $g_{IB}(t>0)=0.1$. The parameters for the effective Hamiltonian are $\overline m_{\mathrm{eff}}/m=1.006$ and $\overline k_{\mathrm{eff}}/k=0.9054$ (see Ref.~\cite{mistakidis2018}). The interaction parameter $g_{II}^i=-0.011g_{BB}$ is identical to the one used in Fig.~\ref{fig:fig2} {\bf a)}. Panel {\bf b)} illustrates the dynamics with $g_{IB}(t>0)=0.3$. The parameters for the effective Hamiltonian, $\overline m_{\mathrm{eff}}/m=1.045$ and $\overline k_{\mathrm{eff}}/k=0.76$, are obtained using the methods of Ref.~\cite{mistakidis2018}. The interaction parameter $g_{II}^i=-0.1g_{BB}$ is identical to the one used in Fig.~\ref{fig:fig2} {\bf b)}.}
  	\label{fig:app_MF} 
\end{figure}
%%%%%%%%%%%%%%%%%%%%%%%%%%%%%%%%%%%%%%%

Having at hand the many-body wavefunction ansatz, we can establish the ML-MCTDHB equations of motion~\cite{mistakidis2017}. To this end, we apply the Dirac-Frenkel variational principle~\cite{Frenkel,Dirac} to the function determined by Eqs.~(\ref{Eq:WF}), (\ref{Eq:SPF}). 
As a result we obtain a set of $D^2$ linear differential equations of motion for the Schmidt coefficients $\lambda_k(t)$ which are coupled to 
$D[$ ${N_B+d_B-1}\choose{d_B-1}$+$\frac{d_I(d_I+1)}{2}$] non-linear integro-differential equations for the species functions and $d_B+d_I$ 
integro-differential equations for the SPFs. 
We note in passing that the ML-MCTDHB method allows us to operate within different approximation 
orders. 
As an example, we retrieve the mean-field equation of motion~\cite{Pethick_book,Kevrekidis_book} of the bosonic mixture when choosing $D=d_B=d_I=1$. 
In this case, the many-body wavefunction ansatz reduces to the well-known mean-field product state  
\begin{equation}
\begin{split}
&\Psi_{MF}(\vec x,\vec y;t) = \varphi_1^I(y_1;t)\varphi_1^I(y_2;t) \prod_{j=1}^{N}\varphi_1^B(x_j;t).  
\label{Eq:MF}
\end{split}
\end{equation} 
Following a variational principle for this ansatz, e.g., due to Dirac and Frenkel~\cite{Frenkel,Dirac}, we arrive at the celebrated system of 
coupled Gross-Pitaevskii equations of motion \cite{Pethick_book,Kevrekidis_book} that govern the dynamics of the bosonic mixture 
\begin{equation}
\begin{split}
 &i\hbar\frac{\partial \varphi_1^B(x;t)}{\partial t}= \bigg[-\frac{\hbar^2}{2m} \frac{\partial^2}{\partial x^2}+\frac{k x^2}{2} 
 +g_{BB}N_B \abs{\varphi_1^B(x;t)}^2+g_{BI} \abs{\varphi_1^I(x;t)}^2 \bigg] \varphi_1^B(x;t), \\  
 &i\hbar \frac{\partial \varphi_1^I(y;t)}{\partial t}= \bigg[-\frac{\hbar^2}{2M} \frac{\partial^2}{\partial y^2}+\frac{k y^2}{2} 
 +g_{II}N_I \abs{\varphi_1^I(y;t)}^2+g_{BI} \abs{\varphi_1^B(y;t)}^2 \bigg] \varphi_1^I(y;t), \\  
 \end{split}
\end{equation} 
where we introduce $N_I=2$ to elucidate the symmetry of the equations. Within the mean-field approximation all particle correlations of the system are neglected, implying that it can be useful only to describe the system at very weak interactions. To illustrate this, we employ the coupled Gross-Pitaevskii equations to study the time evolution of the size of the impurity cloud following a sudden change of $g_{IB}$. We calculate $\int (y_1^2+y_2^2)|\Psi_{MF}|^2\mathrm{d}y_1...\mathrm{d}x_N$ and $2\int y_1^2|\Psi^{(1)}_{MF}|^2\mathrm{d}y_2...\mathrm{d}x_N$, where $\Psi^{(1)}_{MF}$ is the mean-field wave function for a single impurity in a Bose gas~\cite{mistakidis2018}. Following the discussion in the main text, we present in Fig.~\ref{fig:app_MF} the quantity $\langle y^2\rangle_{MF}(t)/\langle y^2\rangle_{MF}(0)$, where $\langle y^2\rangle(t)_{MF}=\int (y_1^2+y_2^2)|\Psi_{MF}|^2\mathrm{d}y_1...\mathrm{d}x_N-\int y_1^2|\Psi^{(1)}_{MF}|^2\mathrm{d}y_2...\mathrm{d}x_N$. We conclude that the parameter $g_{II}^i$ obtained from the ML-MCTDHB calculations agrees well with that for the mean-field calculations only when $g_{IB}(t>0)$ is small, see Fig.~\ref{fig:app_MF} {\bf a)}. Already for $g_{IB}(t>0)=0.3$, beyond-mean-field simulations differ from the mean-field predictions, see Fig.~\ref{fig:app_MF} {\bf b)}.

\section{Convergence and details of the many-body simulations} \label{sec:convergence}

As stated in Appendix~\ref{methodology}, the ML-MCTDHB method is based on the expansion of the many-body wavefunction with respect to 
a time-dependent and variationally optimized basis. 
The underlying Hilbert space truncation is determined by the employed orbital configuration space which we denote as $C=(D;d_B;d_I)$, 
with $D$ and $d_B$, $d_I$ being the number of species and single-particle functions respectively of each species 
[see also Eqs. (\ref{Eq:WF}) and (\ref{Eq:SPF})]. 
For our simulations we utilize a primitive basis based on a sine discrete variable representation including 500 grid points. 
This sine discrete variable representation introduces hard-wall boundary conditions imposed at $x_\pm=\pm40 \mu m$. The presence of the hard-wall boundary does not affect our results, since we do not observe any significant density population beyond $x_{\pm}=\pm23 \mu m$. 

%%%%%%%%%%%%%%%%%%%%%%%%%%%%%%%%%%%%%%%
\begin{figure}
  	\includegraphics[width=0.46\textwidth]{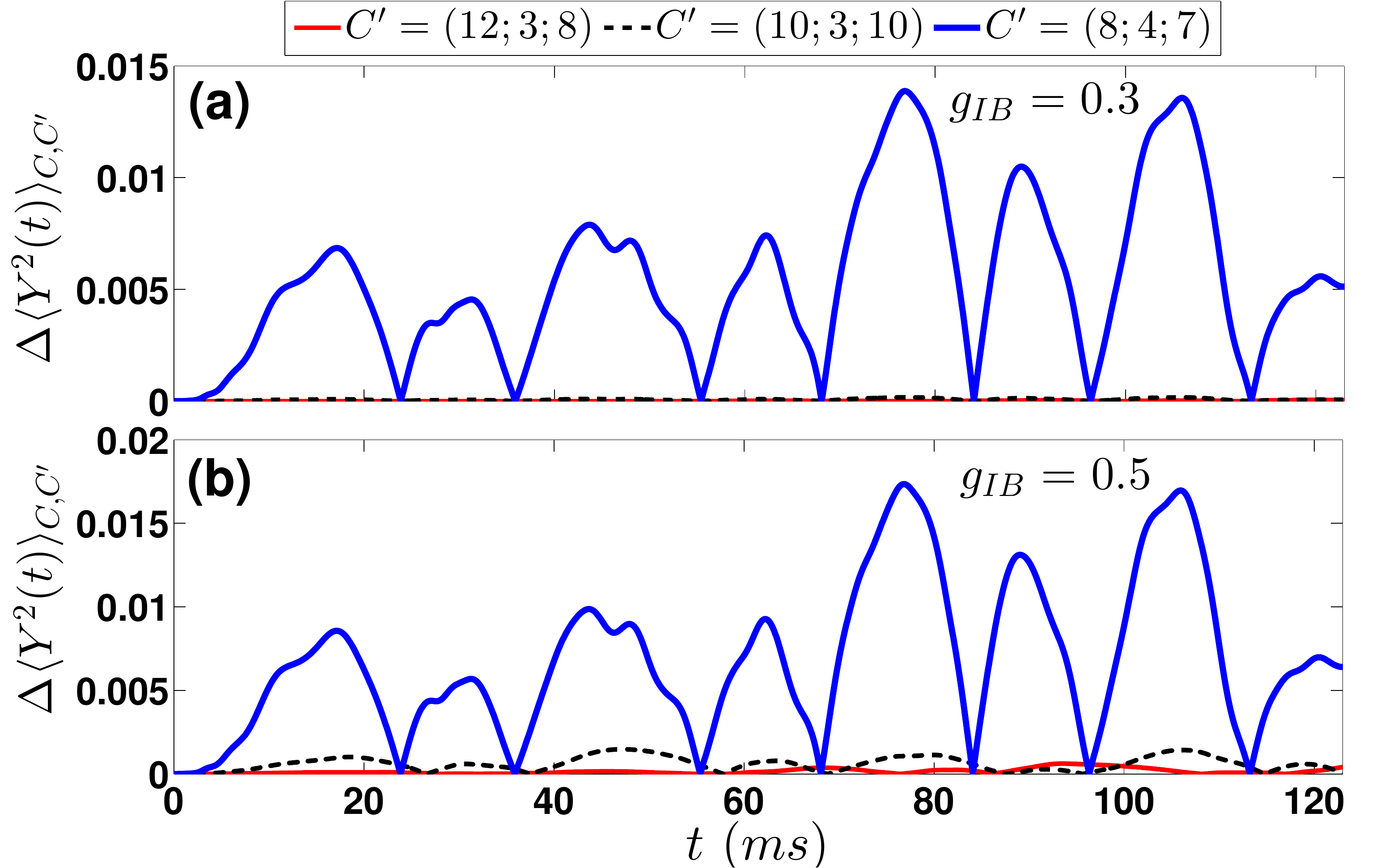}
  	\caption{The time evolution of the deviation $\Delta \braket{Y^2(t)}_{C,C'}$ of the square of the size of the impurity cloud between 
  	the configuration with $C=(10;3;8)$ and other orbital configurations $C'=(D;d_B;d_I)$ (see legend). 
  	The harmonically trapped bosonic mixture contains $N=100$ bosons and two impurity atoms.  
  	The dynamics is induced by a rapid change of the interaction strength from $g_{IB}=0$ to (a) $g_{IB}=0.3$ and (b) $g_{IB}=0.5$.}
  	\label{fig:convergence} 
\end{figure}
%%%%%%%%%%%%%%%%%%%%%%%%%%%%%%%%%%%%%%%

The many-body calculations presented in the main text are based on the orbital configuration $C=(10;3;8)$. We study the convergence of the many-body simulations by varying  the orbital configuration space, $C=(D;d_B;d_I)$. Let us exemplify the convergence of our results for a different number of species and single-particle functions. 
For this investigation we resort to the time-evolution of the variance of the two bosonic impurities, $\braket{Y^2(t)}=\braket{y_1^2+y_2^2}$.   
We study its absolute deviation between the $C=(10;3;8)$ and other orbitals configurations $C'=(D;d_B;d_I)$,
namely, 
\begin{equation}
\begin{split}
\Delta \braket{Y^2(t)}_{C,C'} =\frac{\abs{\braket{Y^2(t)}_C -\braket{Y^2(t)}_{C'}}}{\braket{Y^2(t)}_C}. \label{deviation_variance} 
\end{split}
\end{equation} 
In this expression, $\braket{Y^2(t)}_C$ is calculated for the orbital configuration space $C$.
$\Delta\braket{Y^2(t)}_{C,C'}\to 0$ implies a negligible deviation between $\braket{Y^2(t)}$ 
calculated within the $C$ and $C'$ approximations. 

Figure~\ref{fig:convergence} presents $\Delta\braket{Y^2(t)}_{C,C'}$ for the system considered in the main text with $g_{IB}(t>0)=0.3$ [Fig. \ref{fig:convergence} (a)] and $g_{IB}(t>0)=0.5$ [Fig. \ref{fig:convergence} (b)]. 
A convergence of $\Delta \braket{Y^2(t)}_{C,C'}$ is seen in both cases. 
More specifically, comparing $\Delta \braket{Y^2(t)}_{C,C'}$ between the $C=(10;3;8)$ and $C'=(12;3;8)$ orbital configurations we observe that it acquires values 
below $0.05\%$ ($0.1\%$) for postquench interactions $g_{IB}=0.3$ ($g_{IB}=0.5$) throughout the time evolution. 
Also, the values of $\Delta \braket{Y^2(t)}_{C,C'}$, when $C=(10;3;8)$ and $C'=(8;4;7)$, imply relatively small deviations which become at most of the order of $1.4\%$ ($1.8\%$) 
in the course of the time dynamics for $g_{IB}=0.3$ ($g_{IB}=0.5$). 
The same observations hold also true for the variances of the bosonic gas (not shown here for brevity). 
Finally, we remark that a similar degree of convergence occurs also for the other observables invoked in the present paper,
e.g., the single-particle density of the impurity (not shown for brevity).

\section{The wave function of the effective two-body Hamiltonian}
\label{sec:app_a}

Here we discuss the wave function $\Psi_{\mathrm{rel}}(t)$ that evolves according to the effective Hamiltonian introduced in Eq.~(\ref{eq:eff_ham_rel}) of the main text,
\begin{equation}
\overline H_{\mathrm{rel}}=-\frac{\hbar^2}{2\overline m_{\mathrm{eff}}}\frac{\partial^2}{\partial y^2}+
\frac{\overline k y^2}{2}+\frac{g_{II}^i+g_{II}}{\sqrt{2}} \delta(y).
\label{eq:A1}
\end{equation}
The initial state for the dynamics is the ground state of the one-body Hamiltonian $h(y)=-\frac{1}{2}\frac{\partial^2}{\partial y^2}+\frac{\Omega^2y^2}{2}$ (in this appendix we assume $\hbar=m=1$),
\begin{equation}
\Psi_{\mathrm{rel}}(t=0)=\left(\frac{\Omega}{\pi}\right)^{\frac{1}{4}}e^{-\frac{\Omega y^2}{2}}.
\end{equation}
Note that the Hamiltonian~(\ref{eq:A1}) preserves parity, and that $\Psi_{\mathrm{rel}}(t=0)$ is an even-parity function. Therefore, $\Psi_{\mathrm{rel}}(t)$ must also be of even parity, and we express it as 
\begin{equation}
\Psi_{\mathrm{rel}}(y,t)=\sum_i e^{-iE_i t} a_i \phi_i(y),
\end{equation}
where the set $\{\phi_i(y)\}$ ($\{E_i\}$) consists of all even eigenstates (eigenvalues) of $\overline H_{\mathrm{rel}}$. The expansion coefficients are $a_i=\int \Psi(y,t=0)\phi_i(y)\mathrm{d}y$. The states $\{\phi_i(y)\}$ are known~\cite{busch1998}:
\begin{equation}
\phi_i(y)=N_i e^{-\frac{\overline m_{\mathrm{eff}} \overline \Omega_{\mathrm{eff}} y^2}{2}} 
U\left(-\nu_i,\frac{1}{2},\overline m_{\mathrm{eff}} \overline \Omega_{\mathrm{eff}} y^2\right),
\end{equation}
where $\overline \Omega_{\mathrm{eff}} = \sqrt{ \overline k/\overline m_{\mathrm{eff}}}$, $U$ is Tricomi's confluent hypergeometric function (also known as the confluent hypergeometric function of the second kind)~\cite{abram1972}, 
$\nu_i=\frac{E_i}{2\overline \Omega_{\mathrm{eff}}}-\frac{1}{4}$, and the normalization constant is 
\begin{equation}
N_i=\left(\overline m_{\mathrm{eff}}\overline \Omega_{\mathrm{eff}}\right)^{\frac{1}{4}}\sqrt{\frac{\Gamma(-\nu_i) \Gamma(-\nu_i+\frac{1}{2})}{\pi(\psi(-\nu_i+\frac{1}{2})+\psi(-\nu_i))}}.
\end{equation}
The parameter $\nu_i$ can be found from the equation~\cite{busch1998}
\begin{equation}
\frac{\Gamma(-\nu_i+\frac{1}{2})}{\Gamma(-\nu_i)}=-\frac{g_{II}^i+g_{II}}{2\sqrt{2}\hbar\overline \Omega_{\mathrm{eff}}}\sqrt{\frac{\overline m_{\mathrm{eff}}\overline \Omega_{\mathrm{eff}}}{\hbar}}.
\end{equation}
The expansion coefficients can be expressed in a closed form
\begin{equation}
a_i=\left(\frac{2 c-1}{\pi}\right)^{\frac{1}{4}}
\sqrt{\frac{\Gamma(-\nu_i) \Gamma(-\nu_i+\frac{1}{2})}{\pi(\psi(-\nu_i+\frac{1}{2})+\psi(-\nu_i))}}\left[\frac{\pi(c-1)^{\nu_i}c^{-\nu_i-1/2}}{\Gamma\left(\frac{1}{2}-\nu_i\right)}
+\sqrt{\pi}\frac{3-2\nu_i-c+\left(c+\frac{1}{c}-2\right) \; _2F_1(1,\frac{1}{2}-\nu_i;-\frac{1}{2};\frac{1}{c})}{2\Gamma(2-\nu_i)}\right],
\end{equation}
where $2 c=1+\frac{m\Omega}{\overline m_{\mathrm{eff}} \overline \Omega_{\mathrm{eff}}}$, and $_2F_1$ is the hypergeometric function. 

Once the wave function is derived, we can calculate all observables of interest. For example, the size of the cloud, $\langle y^2\rangle=\int y^2|\Psi(y,t)|^2\mathrm{d}y$, can be computed by truncating the sum
\begin{equation}
\langle y^2\rangle = \sum_{i,j} a_i a_j e^{-i(E_i-E_j)t} A_{ij},
\end{equation}
where $A_{ij}=\int \phi_i y^2 \phi_j\mathrm{d}y$ has an analytic expression~\cite{ebert2016}.
Another observable of interest is the entropy. To calculate it, we need to find the spectral representation of the one-body density matrix for a system of two impurity atoms. This density matrix is derived from the total wave function, which describes both the center-of-mass and the relative dynamics
\begin{equation}
\Psi_{\mathrm{eff}}=\Psi_{\mathrm{CM}}(y_{\mathrm{CM}},t)\Psi_{\mathrm{rel}}(y,t),
\end{equation} 
where the function $\Psi_{\mathrm{CM}}(y_{\mathrm{CM}},t)$ equals to $\phi(y_{\mathrm{CM}},t)$~\cite{mistakidis2018} with
\begin{equation}
\phi(z,t)= \left(\sqrt{\Omega}\frac{\Omega_{\mathrm{eff}} \overline m_{\mathrm{eff}}}{ i \sqrt{\pi}\sin(\Omega_{\mathrm{eff}} t)}\right)^{\frac{1}{2}}\frac{e^{\frac{\overline m_{\mathrm{eff}}z^2\Omega_{\mathrm{eff}}(-i\overline m_{\mathrm{eff}}\Omega_{\mathrm{eff}}-\Omega\cot(\Omega_{\mathrm{eff}}t))}{2 i \Omega+2 \overline m_{\mathrm{eff}} \Omega_{\mathrm{eff}}\cot(\Omega_{\mathrm{eff}}t)}}}{\sqrt{\Omega-i \overline m_{\mathrm{eff}}\Omega_{\mathrm{eff}} \cot(\Omega_{\mathrm{eff}}t)}}.
\end{equation}
The density matrix can be written as 
\begin{equation}
\rho^{(I)}_{\mathrm{eff}}=\sum_{i,i'} \beta_{i,i'}(t)f_i^*(y,t)f_{i'}(y',t).
\end{equation}
In this equation, $\{f_i\}$ is an orthonormal set of functions constructed such that the function $f_{i>0}$ is orthogonal to $\phi$, 
\begin{equation}
f_{i}(y,t)=\frac{1}{\sqrt{2^n n!}}\left(\frac{\delta(t)}{\pi}\right)^{1/4}e^{-\frac{\delta(t)y^2+i\Delta(t)y^2}{2}}H_{n}\left(\sqrt{\delta(t)}y\right),
\end{equation} 
where $H_n$ is the $n$th Hermite polynomial. The functions $\delta(t)$ and $\Delta(t)$ are expressed as
\begin{align}
\delta(t)=\frac{\overline m_{\mathrm{eff}}^2\Omega^2_{\mathrm{eff}}\Omega(1+\cot^2(\Omega_{\mathrm{eff}}t))}{\Omega^2+\overline m_{\mathrm{eff}}^2\Omega^2_{\mathrm{eff}}\cot^2(\Omega_{\mathrm{eff}}t)}, \qquad
\Delta(t)=\frac{\overline m_{\mathrm{eff}}\Omega_{\mathrm{eff}}\cot(\Omega_{\mathrm{eff}}t)(\overline m^2_{\mathrm{eff}}\Omega^2_{\mathrm{eff}}-\Omega^2)}{\Omega^2+\overline m_{\mathrm{eff}}^2\Omega^2_{\mathrm{eff}}\cot^2(\Omega_{\mathrm{eff}}t)}.
\end{align}
At every time instant $t$, the parameter $\delta(t)$ defines  a time-independent harmonic oscillator whose eigenstates are given by $f_i(t)$.
The expansion coefficients of $\rho^{(I)}_{\mathrm{eff}}$ read $\beta_{i,i'}(t)=\sum_j\alpha_{i,j}^*(t)\alpha_{i',j}(t)$, where
\begin{equation}
\alpha_{i,j}=\int \Psi_{\mathrm{eff}}(y,Y) f_{i}(y_1,t)^*f_{j}(y_2,t)^*\mathrm{d}y_1\mathrm{d}y_2,
\end{equation}
here $y=(y_1-y_2)/\sqrt{2}$ and $Y=(y_1+y_2)/\sqrt{2}$. 
The expression for $\alpha_{i,j}$ can be simplified as follows
\begin{align}
\alpha_{i,j}&=\sqrt{\frac{(i+j)!}{2^{i+j} i!j!}}\left(\sqrt{\Omega}\frac{\Omega_{\mathrm{eff}} \overline m_{\mathrm{eff}}}{ i \sin(\Omega_{\mathrm{eff}} t)}\right)^{\frac{1}{2}}\frac{\delta(t)^{-1/4}}{\sqrt{\Omega-i \overline m_{\mathrm{eff}}\Omega_{\mathrm{eff}} \cot(\Omega_{\mathrm{eff}}t)}}\int \Psi_{\mathrm{rel}}(y,t)f_{i+j}^{*}(y,t)\mathrm{d}y\nonumber \\
&=\sqrt{\frac{(i+j)!}{2^{i+j} i!j!}}\left(\sqrt{\Omega}\frac{\Omega_{\mathrm{eff}} \overline m_{\mathrm{eff}}}{ i \sin(\Omega_{\mathrm{eff}} t)}\right)^{\frac{1}{2}}\frac{\delta(t)^{-1/4}}{\sqrt{\Omega-i \overline m_{\mathrm{eff}}\Omega_{\mathrm{eff}} \cot(\Omega_{\mathrm{eff}}t)}}\sum_k e^{-i\frac{E_k t}{\hbar}}a_k \int \phi_{k}(y)f^*_{i+j}(y,t)\mathrm{d}y.
\end{align}
Note that $\alpha_{i,j}=0$ if $i+j$ is an odd number. The same is true for $\beta_{i,j}$. To obtain the spectral representation of the density matrix, which allows us to calculate the entropy, we find and diagonalize the matrix $\beta_{i,i'}$.

 \end{document}